\documentclass[preprint,prb,amsmath]{revtex4-2}

\usepackage{fontspec} 
\usepackage{amsmath,amssymb,amsfonts,xcolor,graphicx,cases}

\makeatletter
\let\old@makecaption=\@makecaption
\usepackage{subcaption}
\let\@makecaption=\old@makecaption
\makeatother
\usepackage{blindtext}

\usepackage{hyperref}

\def\be{\begin{equation}}
\def\ee{\end{equation}}
\def\bea{\begin{eqnarray}}
\def\eea{\end{eqnarray}}
\def\bsn{\begin{subnumcases}}
\def\esn{\end{subnumcases}}

\def\eq#1{(\ref{#1})}
\def\bm{\begin{pmatrix}}
\def\em{\end{pmatrix}}

\begin{document}

\title{Kink-kink correlations in nonlinear quenches across a quantum critical point}
\author{Lakshita Jindal}
\email{lakshita@jncasr.ac.in}
\author{Kavita Jain} 
\email{jain@jncasr.ac.in}
\affiliation{Theoretical Sciences Unit, \\Jawaharlal Nehru Centre for Advanced Scientific Research, \\Bangalore 560064, India}

\date{\today}
\begin{abstract}
When a quantum system exhibiting a second order phase transition is quenched across the critical point in large but finite time, the dynamics are not adiabatic in the critical region and the Kibble-Zurek (KZ) mechanism provides a framework to determine local observables such as the mean defect density. 
However, to find higher-point functions,  one has to go beyond the KZ paradigm as shown in recent works on one-dimensional transverse field Ising model (TFIM) following a linear quench. It has been found that 
(i) besides the KZ scale, the quench dynamics  depend on another length scale that arises due to the finite phase difference between the low energy modes, and (ii) contrary to the expectations based on the KZ mechanism, in general, the correlation functions do not decay exponentially with distance. Motivated by these results for the linear quench, we are interested in understanding if these properties are universal, and 
consider the $1$D TFIM when  the transverse field varies algebraically in the vicinity of the critical field. We focus on the equal-time, longitudinal kink-kink correlation function at the end of the quench from the paramagnetic to the ferromagnetic phase, and find that (i) the correlator depends only on the KZ length for superlinear quenches, otherwise an additional dephasing length is required to describe it, and (ii) the dephased correlator decays 
as a compressed exponential with an exponent that changes continuously with the quench exponent. Our results are obtained using an adiabatic perturbation theory, analytical arguments and exact numerical integration of the relevant equations. 
\end{abstract}

\maketitle

\clearpage
\section{Introduction} 
\label{sec_intro}

Understanding the nonequilibrium dynamics of quantum systems, particularly across quantum phase transitions, is important to gain valuable insights into fundamental physics \cite{Sachdev:2011} and emerging quantum technologies \cite{Georgescu:2014,Albash:2018,Scully:2025}. As an example, consider adiabatic quantum computing  \cite{Albash:2018} in which starting from an easy-to-prepare ground state of a quantum system, the Hamiltonian is varied slowly so as to arrive at the ground state of a Hamiltonian that encodes the solution of a complex problem. However, if a phase transition occurs during such annealing dynamics, diabatic transitions can occur in the critical region. 
Then it is natural to ask how the loss of adiabaticity depends on the rate at which the system is driven across the critical point, how the defect density resulting from excitations depends on the quench rate,
 and how the system evolves towards its final ground state. The Kibble-Zurek (KZ) argument \cite{Kibble:1976,Zurek:1985} described below provides a useful framework to address some of these questions.

When a second order phase transition occurs in the ground state of an infinitely large quantum system by tuning a parameter $g$, a perturbation in the ground state relaxes on a time scale $T_{relx}\sim \xi^z$, where the correlation length $\xi \sim |g-g_c|^{-\nu}$ in the neighborhood of the critical point $g_c$. If now such a system is slowly quenched to or across the critical point in time $\tau \gg 1$, then the inverse rate of change of energy, $\delta_g=\frac{1}{E} \frac{dE}{dt}$ due to the change in parameter $g$ provides another characteristic time scale in the system. Far from the critical point where $T_{relx} \ll \delta^{-1}_g$, the system can relax to the ground state of the Hamiltonian at the instantaneous $g$ and  the time evolution is adiabatic. But as the critical point is approached, the relaxation time grows, and there exists a time scale ${\hat t}$ where $T_{relx}({\hat t}) \sim \delta^{-1}_g({\hat t})$. 
As a result, the system can not evolve adiabatically during the impulse regime, ${\hat t} < t < 2 t_c-{\hat t}$, where $g_c=g(t_c)$. But beyond this regime where the relaxation time is again small, the adiabatic evolution resumes \cite{Kibble:1976,Zurek:1985}. 
Then, if the parameter $g$ varies algebraically in the critical region, $\left|g(t)-g_c\right| \sim \left|\frac{t_c-t}{t_c}\right|^\omega$ where $t_c$ is of order $\tau$, and the ground state energy is also algebraic in the deviation $|g(t)-g_c|$, on matching the characteristic time scales, it follows that   $t_c -{\hat t} \sim {\hat \xi}^z \sim \left|\frac{t_c-{\hat t}}{t_c}\right|^{-\omega \nu z}$, or $|t_c-{\hat t}| \sim t_c^{\frac{\omega \nu z}{1+\omega \nu z}}$.

In the adiabatic-impulse-adiabatic (AIA) picture described above, it is also assumed that during the impulse regime, the time evolution freezes.
This  argument has been further elaborated \cite{Damski:2005,Damski:2006,Sadhukhan:2020,Suzuki:2025}, and its scaling predictions for local observables have been verified  in numerous classical \cite{Jelic:2011, Priyanka:2016, Ricateau:2018, Jeong:2020, Sun:2020, Priyanka:2021, Jeong:2022, Jindal:2024, Tang:2025, Shu:2026} and quantum \cite{Sen:2008, Kolodrubetz:2012b, Puskarov:2016, Campo:2018, Zamora:2020, Schmitt:2022, Kou:2023, Suzuki:2024, Jindal:2025, Soto:2026} models as well as experiments 
in various condensed matter systems \cite{Maegochi:2022, Du:2023, Wang:2025}, superfluids \cite{Ko:2019,Lee:2024}, ultracold atomic systems \cite{Labeyrie:2016, Cheng:2024}, colloidal systems \cite{Deutschlander:2015, Libal:2020}, semiconductors \cite{Wang:2014} and quantum annealers \cite{Bando:2020,King:2022,Miessen:2024}. For additional examples on quantum systems, one may refer to these reviews \cite{Dziarmaga:2010,Polkovnikov:2011,Dutta:2015, Campo:2014}. While intuitively appealing, the AIA picture is quantitatively inaccurate \cite{Damski:2005,Damski:2006}, and much progress in our understanding of the slow quench dynamics has been made via the transverse field Ising model (TFIM) which can be solved exactly for linear quenches \cite{Dziarmaga:2005} and was recently implemented in a quantum annealer \cite{King:2022}.

Although much is known for TFIM following a linear quench, far fewer results are available for nonlinear quenches and  are limited to local observables such as  the excitation probability and the mean defect density \cite{Garanin:2002,Sen:2008,Barankov:2008}. 
Recent works on linear quenches have shown that, contrary to the expectations based on the AIA picture \cite{Damski:2005,Damski:2006}, at the end of the quench, the transverse spin-spin correlation function \cite{Cherng:2006,Cincio:2007} and the kink-kink correlation function in either direction \cite{Roychowdhury:2021} decay as a Gaussian with distance, and 
these correlation functions depend not only on the KZ length, ${\hat \xi} \sim \sqrt{\tau}$, but also on a longer dephasing length, $\ell \sim \sqrt{\tau} \ln \tau$ \cite{Cincio:2007,Nowak:2021,Dziarmaga:2022}; however, no simple interpretation or understanding of the latter scale was provided. Although the dephasing length is briefly discussed in \cite{Kolodrubetz:2012b} for linear and nonlinear quenches, no connection was made to the prior work in \cite{Cincio:2007} and the results for nonlinear quenches were not tested.

Here we are interested in understanding how the aforementioned results for correlation functions are affected when the quench is nonlinear. We consider a generic protocol in which close to the critical point, the transverse field varies algebraically, and focus on a kink-kink correlation function at the end of the quench from the paramagnetic to the ferromagnetic phase. 
Although it does not appear possible to solve the nonlinear quench problem exactly, for slow quenches, an adiabatic perturbation theory  \cite{Garanin:2002,Polkovnikov:2005,Grandi:2010} is used to obtain an analytical expression for the dephased correlator, and we find that it decays super-exponentially with distance for any quench exponent. 
Generalizing an argument in \cite{Kolodrubetz:2012b}, we also find that for superlinear quenches, a single length scale, namely, the KZ length is sufficient to describe the kink-kink correlator, but for sublinear and linear quenches, the dephasing length scale is also required. We also discuss quench protocols for which the dephasing is optimized. All the analytical results are verified or compared with those obtained by numerically integrating the relevant equations.


\section{Model} 
\label{sec_model}

We consider a TFIM on a ring with $N$ sites defined by the Hamiltonian,
\be
{H(t)} = -J \sum_{j=1}^N {\sigma}^z_j {\sigma}^z_{j+1} - g(t)\sum_{j=1}^N {\sigma}^x_j
\label{hamil}
\ee
where ${\sigma}$'s are Pauli matrices, $J>0$ is the ferromagnetic coupling between the spins, $g(t) \geq 0$ is the time-dependent magnetic field applied in the transverse direction, and ${\sigma}_{N+1}={\sigma}_1$. A more general Hamiltonian in which both $J$ and $g$ are time-dependent can also be considered, but a suitable transformation can be made to obtain a time-independent coupling \cite{Ashhab:2022}. 

We first summarize the pertinent results when the transverse field is constant in time \cite{Sachdev:2011,Mbeng:2024}. At zero temperature, in the thermodynamic limit $(N\to \infty)$, this model exhibits a phase transition at the critical field $g_c=1$ between a paramagnetic ($g>1$) and ferromagnetic ($g<1$) phase. The Hamiltonian (\ref{hamil}) can be diagonalized by mapping it to a spinless fermionic Hamiltonian via Jordan-Wigner transformation, $\sigma_j^x = 1-2c_j^\dagger c_j, \sigma_j^z = (c_j^\dagger + c_j) \prod_{i=1}^{j-1} (1-2c_i^\dagger c_i) $ where $c_j$'s are the fermionic operators. In momentum space, one then obtains 
\be
H =  \sum_{k>0} 2 J (g-\cos k)(c_k^\dagger c_k -c_{-k}c_{-k}^\dagger) - 2i J \sin k(c_{k}^\dagger c_{-k}^\dagger -c_{-k}c_k )
\label{hamfou}
\ee
where $c_k = \frac{1}{\sqrt{N}}\sum_{j=1}^N e^{-ikj}c_j$ satisfy anticommutation relations, $\{c_k,c_p\}=\{c_k^\dagger,c_p^\dagger \}=0, \{c_k^\dagger,c_p \} =\delta_{k,p}$, and the momenta $k=\pm \frac{(2 m+1) \pi}{N}, m=0, 1, ...,\frac{N-2}{2}$ for anti-periodic boundary conditions, $c_{N+1}=-c_1$ (assuming $N$ is even). The summand on the RHS of Eq. \eq{hamfou},  which we denote by $H_k$, can be diagonalized through a Bogoliubov transformation to fermionic annihilation operator, $\gamma_k=v_{k,eq} c_k - u_{k,eq} c_{-k}^\dagger$ 
to yield $H_k=\epsilon_k (\gamma_k^\dagger \gamma_k- \gamma_{-k} \gamma_{-k}^\dagger), k > 0$ where, 
\bea
\epsilon_k &=&  2 J \sqrt{1+g^2-2g\cos k} \label{disrel} \\
u_{k,eq} &=& \frac{2 i J \sin k}{\sqrt{4 J^2 \sin^2 k+(\epsilon_k+2 J (g -\cos k))^2}} \label{ueq} \\
v_{k,eq}&=& \frac{\epsilon_k+2 J (g -\cos k)}{\sqrt{4 J^2 \sin^2 k+(\epsilon_k+2 J (g -\cos k))^2}}  \label{veq}
\eea
with $|u_{k,eq}|^2+|v_{k,eq}|^2=1$. For the two-level system defined by the Hamiltonian $H_k$, the ground state and the excited state  with respective energies $-\epsilon_k$ and $+\epsilon_k$ are given by 
\bea
| \phi^{(0)}_{k,eq} \rangle &=& [v_{k,eq}+u_{k,eq} c_k^\dagger c_{-k}^\dagger] |0 \rangle \label{phi0eq}\\ 
| \phi^{(1)}_{k,eq} \rangle &=&[u_{k,eq}+v_{k,eq} c_k^\dagger c_{-k}^\dagger] |0 \rangle \label{phi1eq}
  \eea
where $|0\rangle$ is the vaccum state with $c_{\pm k} |0\rangle=0$. 
The Hamiltonian (\ref{hamfou}) thus describes  an ensemble of two-level systems with  Hamitonian $H_k$ and its ground state 
is given by the tensor product, $|\emptyset_{eq} \rangle =  \otimes_{k > 0} | \phi^{(0)}_{k,eq} \rangle$. 

For the time-dependent Hamiltonian \eq{hamil}, in the Heisenberg picture, the operator $c_k(t) = v_k(t) \gamma_k-u_{k}(t) \gamma^{\dagger}_{-k}$ with time-dependent coefficients that evolve as \cite{Dziarmaga:2005,Dziarmaga:2010}, 
\be
\label{sm_ukvk}
i \hbar \frac{d}{dt} \bm u_k \\ v_k \em= \bm -2 J (g(t) -\cos k) & 2 i J \sin k \\  -2 i J \sin k & 2 J (g(t) -\cos k) \em \bm u_k \\ v_k \em
\ee
Then the time-dependent eigenfunctions of $H_k(t)$ are obtained on replacing $u_{k,eq}$ and $v_{k,eq}$ by $u_k(t)$ and $v_k(t)$, respectively, in Eqs. \eq{phi0eq} and \eq{phi1eq}. 
For the kink-kink correlator, as discussed in Sec.~\ref{qties}, it is convenient to work with dual variables defined as $\mu_i^x = \sigma_i^z \sigma_{i+1}^z, \mu_{j}^z = \prod_{i=1}^{j} \sigma_i^x$ in terms of which one obtains the Hamiltonian \eq{hamil} but with  $J$ and $g$ replaced by $J'=Jg$ and $g'=1/g$, respectively, and the corresponding coefficients $u'_k, v'_k$ now obey
\be
\label{sm_ukpvkp}
i \hbar \frac{d}{dt} \bm u'_k \\ v'_k \em= \bm -2 J (1-g(t) \cos k) & 2 i J g(t) \sin k \\  -2 i J g(t) \sin k & 2 J (1-g(t) \cos k) \em \bm u'_k \\ v'_k \em
\ee
A unitary transformation to variables $\tilde u_k, \tilde v_k$ given by
\bea
{\tilde u}_k &=& \cos(k/2) {u}'_k +i \sin(k/2) {v}'_k \label{uutilde} \\
{\tilde v}_k &=& i \sin(k/2) {u}'_k + \cos(k/2) {v}'_k  \label{vvtilde}
\eea
renders the off-diagonal elements in the matrix on the RHS of Eq. \eq{sm_ukpvkp} to be time-independent where $\tilde u_k$ and $\tilde v_k$, respectively, obey the same equations as $-v_k$ and $u_k$   \cite{Jindal:2025}:
\be
\label{sm_uktvkt}
i \hbar \frac{d}{dt} \bm {\tilde u}_k \\ {\tilde v}_k \em= \bm 2 J (g(t) -\cos k) & 2 i J \sin k \\  -2 i J \sin k & -2 J (g(t) -\cos k) \em \bm {\tilde u}_k \\ {\tilde v}_k \em
\ee
In the following discussion, we will use either Eq.  \eq{sm_ukpvkp} or \eq{sm_uktvkt}, as per convenience.

\section{Quench protocol}
\label{sec_proto}

We consider a quench protocol that can not be linearized in the vicinity of the critical point and 
in which the transverse field varies as a power law close to $g_c$ with an exponent $\omega > 0$. Specifically, we assume that the system is quenched from $g_i$ to $g_f$ in time $\tau$ with 
\bsn
{g(t)= \label{nonlinprot} }
        g_c+(g_i-g_c) \left(\frac{t_c-t}{t_c}\right)^\omega, & $0 \leq t < t_c $ \\
        g_c-(g_c-g_f) \left(\frac{t-t_c}{\tau-t_c}\right)^\omega, & $t_c \leq t \leq \tau$   
\esn
where $g(t_c)=g_c$. Note that the transverse field $g(t)$ is a continuous function of time for finite $\omega$, but  for $\omega < 1$, its derivative w.r.t. time diverges at the critical point. We focus on a class of protocols where the system is initially prepared in the ground state at $g_i > 1$ and quenched to $g_f=0$, where the time $t_c$ to reach the critical point is of order $\tau$. 

\begin{figure}[t]
     \centering
     \begin{subfigure}{0.49\textwidth}
         \centering
         \includegraphics[width=1.0\textwidth]{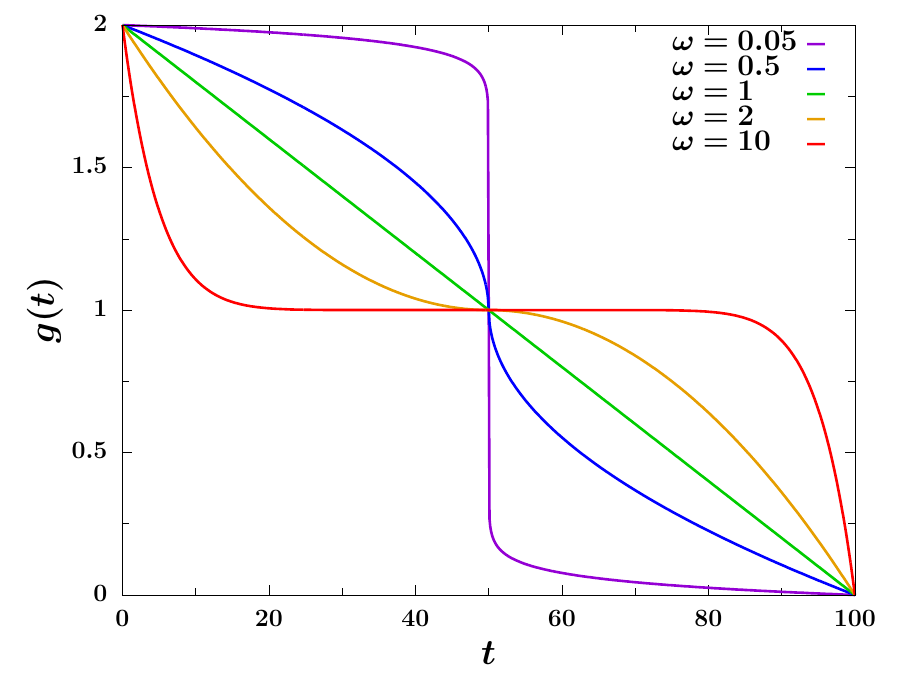}
                 \caption{}
        \label{prot_a}
     \end{subfigure}
     \begin{subfigure}{0.49\textwidth}
         \centering
         \includegraphics[width=1.0\textwidth]{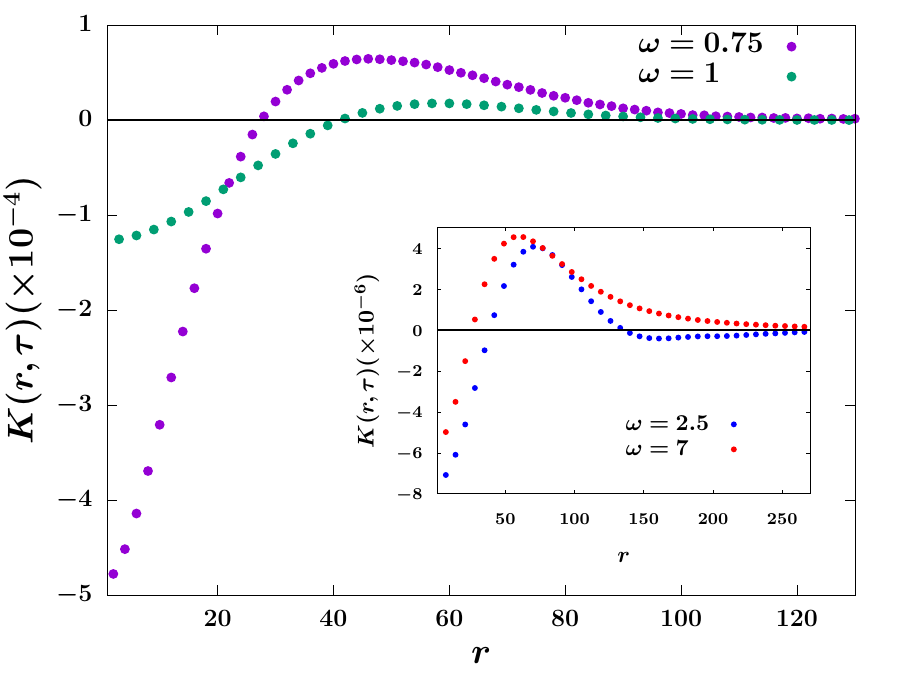}
         \caption{}
        \label{prot_b}
     \end{subfigure}
     \caption{(a) The nonlinear quench protocol (\ref{nonlinprot}) with parameters $g_i=2, g_f=0$, $\tau=10^2$, $t_c=0.5\tau$. (b) The longitudinal kink-kink correlator (\ref{sm_corrdef}) at the end of the quench as a function of distance between kinks for various values of $\omega$; here, 
 $g_i=10, g_f=0$, $\tau=10^3$, $t_c=0.9\tau$.}
\label{fig_prot}
\end{figure}

The protocol \eq{nonlinprot} is shown for some representative values of $\omega$ in Fig.~\ref{prot_a}. For $\omega \to 0$, the transverse field changes rapidly  at $t=t_c$ from $g_i$ to $g_f$, 
while $\omega \to \infty$ corresponds to an instantaneous quench from $g_i$ to $g_c$ followed by free evolution at $g_c$ for time $\tau$ \cite{Barouch:1970,Calabrese:2011,Calabrese:2012}.  In most of the previous work, the system is quenched linearly from $g_i \to \infty$ to $g_f=0$ with $g(t)=-\frac{t}{\tau}, -\infty < t < 0$ \cite{Dziarmaga:2010,Polkovnikov:2011,Campo:2014,Dutta:2015}; this protocol corresponds to $\omega=1$ with the slope $dg/dt$ above and below $t_c$ to be the same, that is, $\frac{t_c}{\tau-t_c}=g_i-1$ in Eq. \eq{nonlinprot}. In the above protocol, the time spent in either phase can also be varied by changing $t_c$; although, as expected, this parameter does not  make a qualitative difference, it affects the quantities of interest quantitatively as discussed in the following sections.

\section{Equal-time longitudinal kink-kink correlator}
\label{qties}

Since a kink between sites $j$ and $j+1$ along the longitudinal direction  is given by the operator $\frac{1- \sigma_j^z\sigma_{j+1}^z}{2}$, due to translational invariance, the equal-time longitudinal  (connected) kink-kink correlation function is defined as 
\bea
K(r, t) &=& \frac{1}{4}\left[\langle \left( 1- \sigma_j^z\sigma_{j+1}^z\right) \left( 1- \sigma_{j+r}^z\sigma_{j+r+1}^z\right) \rangle -  \langle 1- \sigma_j^z\sigma_{j+1}^z \rangle  \langle 1- \sigma_{j+r}^z\sigma_{j+r+1}^z \rangle \right] \label{Kdef1}\\
&=& \frac{1}{4}\left(\langle \sigma_j^z \sigma_{j+1}^z \sigma_{j+r}^z \sigma_{j+r+1}^z \rangle - \langle \sigma_j^z \sigma_{j+1}^z \rangle \langle \sigma_{j+r}^z \sigma_{j+r+1}^z \rangle \right)
\eea
where, the angular brackets denote the average w.r.t. the time-dependent ground state. In terms of dual variables defined in Sec.~\ref{sec_model}, the above equation simplifies to give
\bea
K(r, t) &=& 
\langle c_j^\dagger c_j c_{j+r}^\dagger c_{j+r} \rangle -  \langle c_j^\dagger c_j \rangle \langle c_{j+r}^\dagger c_{j+r} \rangle \\
&=& \langle c_j^\dagger c_{j+r} \rangle \langle c_j c_{j+r}^\dagger  \rangle - \langle c_j^\dagger c_{j+r}^\dagger \rangle \langle c_j c_{j+r} \rangle 
\eea
where the last expression follows on using the Wick's theorem, and the ground state vector $|\emptyset \rangle = \otimes_{k > 0} [v'_{k}(t)+u'_{k}(t) c_k^\dagger c_{-k}^\dagger] |0 \rangle $ with $u_k', v_k'$ being the solution of Eq. \eq{sm_ukpvkp}.
Then, in momentum space and on taking $N \to \infty$, one finally obtains \cite{Roychowdhury:2021,Nowak:2021,Dziarmaga:2022}
\bea
K(r,t) &=& \frac{1}{\pi^2} \left[\int_0^\pi dk \; |u'_k|^2 \cos (k r)  \int_0^\pi dp\; |v'_{p}|^2 \cos (p r)+\bigg|\int_0^\pi dk \; v'_k u_k'^{*} \sin(k r)\bigg|^2  \right] \label{sm_corrdef}
\eea

As $\left(\frac{1-\sigma^z_j \sigma^z_{j+1}}{2}\right)^2 = \frac{1-\sigma^z_j \sigma^z_{j+1}}{2}$, it follows from the definition \eq{Kdef1}  that $K(0,t)=\rho_d(t) (1-\rho_d(t))$ where, due to above equation, the longitudinal mean defect density is given by
\bea
\rho_d(t)&=& \frac{1}{\pi} \int_0^\pi dk \; |u'_k|^2 \label{rhoddef}
\eea
For slow quenches, the kinks are produced in the impulse regime (${\hat t} < t < 2 t_c-{\hat t}$) where the energy gap is small, and $|u'_k|^2$ is interpreted as the excitation probability of a quasiparticle with momentum $k$. 
But as the defect density does not change substantially in the adiabatic regime in the ferromagnetic phase (see, for e.g., Fig. 1a of \cite{Jindal:2025}), the excitation probability at the end of the quench is essentially the same as in the impulse regime, and will be discussed in the following section.

For $r > 0$, Eq. \eq{sm_corrdef} simplifies to give 
\bea
K(r, t) &=& \left| \frac{1}{\pi}\int_0^\pi dk \sin(kr)\; u_k'^{*}  v'_k \right|^2  - \left|\frac{1}{\pi} \int_0^\pi dk \cos(kr)\; |u'_k|^2 \right|^2 
\label{sm_corrapprox}
\eea
which is a sum of two terms: the first term on the RHS of the above equation (henceforth denoted by $K_{\textrm{off}}(r, t)$) depends on the off-diagonal elements of the density matrix of the two-level system with Hamiltonian $H_k$ (refer to Sec.~\ref{sec_model})  and {captures the effects of quantum interference}, while the second term (denoted by $K_{\textrm{on}}(r, t)$) describes the dephased correlator and is related to the excitation probability. We mention in passing that although here we  focus on the longitudinal kink-kink correlation function, due to duality, our results are expected to hold for transverse spin-spin correlation function also \cite{Cherng:2006,Cincio:2007}.

In the following sections, we study the kink-kink correlator analytically using an adiabatic perturbation theory and by numerically integrating Eq. \eq{sm_ukpvkp}.  We assume that the system is initially in the ground state at $g_i$ so that  $u'_k(0)=u'_{k,eq}, v'_k(0)=v'_{k,eq}$, which are obtained on replacing $J \to Jg_i$ and $g \to 1/g_i$ in Eqs.~(\ref{ueq}) and (\ref{veq}), respectively. In the numerical calculations, $N=10^4, J=1$ and $\hbar=1$ are fixed, and all the quantities are measured at the end of the quench. 
In a finite system, the relevant length scales are the chain length $N$; the KZ correlation length ${\hat \xi} \sim \tau^{\frac{\omega}{1+\omega}}$ \cite{Sen:2008} which is consistent with that obtained in Sec.~\ref{sec_intro} on using that the critical exponents, $z=\nu=1$ for TFIM \cite{Sachdev:2011}; and the dephasing length $\ell$ given by Eq.  \eq{ellform} below. To avoid finite-size effects, we require $N \gg \max\{{\hat \xi}, \ell\}$; here, as we typically used $\tau \sim 10^3-10^4$, this condition is satisfied.

 The numerical results for the kink-kink correlator (\ref{sm_corrdef}) at the end of quench are shown in Fig.~\ref{prot_b} for various $\omega$. We find that at short distances, the correlation function is negative which means that the kinks are anti-correlated and `repel' each other, but at intermediate distances, the presence or absence of kinks has positive correlation and finally, at longer distances, the  correlations vanish. As the negative correlations arise due to $K_{\text{on}}$ while the origin of positive correlations lies in $K_{\text{off}}$, 
in the following sections, we analyze the two terms in Eq. (\ref{sm_corrapprox}) separately, and find the length scales associated with them.

\section{Excitation probability} 
\label{excit}

As discussed in the preceding section, to find the dephased correlator, we require the excitation probability $p_k \equiv |u'_k|^2$ where, $u_k'$ obeys Eq. \eq{sm_ukpvkp}. But as the low energy modes are excited in the impulse regime, $|u'_k|^2 \stackrel{k \to 0}{\approx} |{\tilde u}_k|^2$ due to Eq. \eq{uutilde}. 
The coefficient ${\tilde u}_k$ obeys Eq. \eq{sm_uktvkt} which defines a two-level Landau-Zener problem (for a recent review, see \cite{Ivakhnenko:2023}), with Hamiltonian $H_k(t)$ and eigenvalues $\pm \epsilon_k(t)$ given by Eq.  \eq{disrel} for time-dependent $g$. 

\begin{figure}[t]
     \centering
     \begin{subfigure}{0.49\textwidth}
         \centering
         \includegraphics[width=1.0\textwidth]{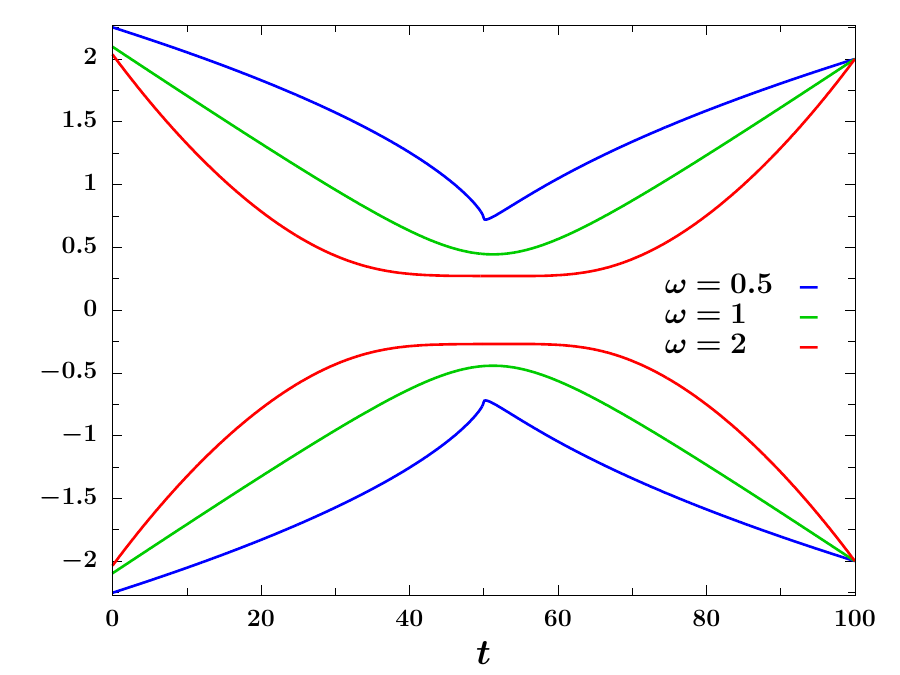}
         \caption{}
        \label{level_a}
     \end{subfigure}
     \begin{subfigure}{0.49\textwidth}
         \centering
         \includegraphics[width=1.0\textwidth]{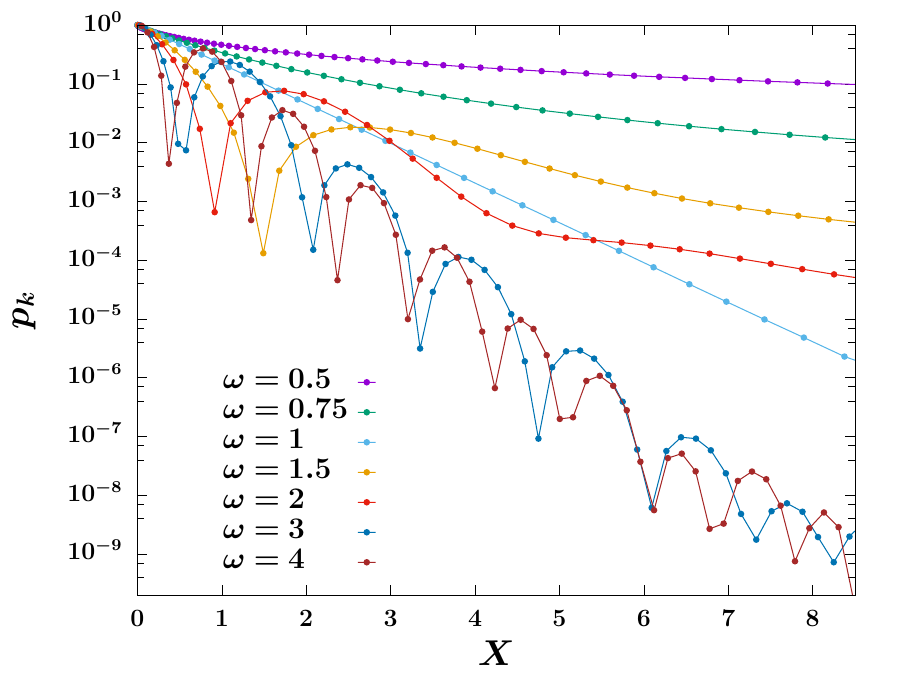}
         \caption{}
        \label{level_b}
     \end{subfigure}
     \caption{(a) Energy level diagram  obtained on using Eq. (\ref{nonlinprot}) in Eq. (\ref{disrel}) with the curves below (above) the $x$-axis showing $-\epsilon_k(t)$ ($+\epsilon_k(t)$) for $k=(\frac{5}{\tau})^{\frac{\omega}{1+\omega}}, \tau=10^2, g_i=2, g_f=0, t_c=0.5\tau$. (b) Excitation probability as a function of $X=\frac{2 \tau k^{\frac{1+\omega}{\omega}}}{\omega}$ for various $\omega$  with $ \tau=10^3, g_i=2, g_f=0, t_c=0.5\tau$ in the quench protocol (\ref{nonlinprot}) to  show that it decreases with increasing $\omega$; here, the solid lines are guide to the eye.}
     \label{fig_level}
\end{figure}

The energy level structure for various $\omega$  is shown in Fig.~\ref{level_a}; as $|\frac{d \epsilon_k}{dt}| \stackrel{k \to 0}{\to} 2 J  |\frac{d g}{dt}|$, for slow quenches, the change in energy in the neighborhood of $g_c$ is small for $\omega \geq 1$ and therefore, one expects that the
ground state can evolve adiabatically. But for  $\omega < 1$, as $\frac{dg}{dt} \propto |t-t_c|^{\omega-1}$, the chance of diabatic transitions is high. 
These qualitative expectations are indeed in accordance with the data shown in Fig.~\ref{level_b}. Also, for large $\omega$, 
as the energy gap remains nearly constant for most of the time (see, Fig.~\ref{level_a}), multiple transitions between the ground and the excited state can occur resulting in the oscillatory behavior of $p_k$  \cite{Garanin:2002}. 


Except for linear quenches \cite{Dziarmaga:2005}, it does not appear possible to obtain $p_k$ exactly, but it can be estimated using an adiabatic perturbation theory  \cite{Garanin:2002,Polkovnikov:2005,Grandi:2010}. 
For a slowly varying Hamiltonian $H_k(t)$ for which $i \hbar \frac{\partial |{\tilde \psi}\rangle}{\partial t}=H_k(t) |{\tilde \psi}\rangle$,  the state vector $|{\tilde \psi}(t)\rangle$ 
 can be expanded in an adiabatic basis with time-dependent coefficients; here, for Eq. \eq{sm_uktvkt}, we have  $|{\tilde \psi}(t)\rangle=\sum_{n=0,1} a_n(t) e^{-i \int_0^t E_n(t') dt'} | \phi^{(n)}_{k,eq}(t) \rangle$ where, $H_k(t) | \phi^{(n)}_{k,eq}(t) \rangle=E_n(t) | \phi^{(n)}_{k,eq}(t) \rangle$ with $E_1=-E_0=\epsilon_k$, and the energy $\epsilon_k$ is given by  Eq. \eq{disrel}. Then, using that for large quench times, the probability of diabatic transition is small ($a_0 \approx 1, a_1 \approx 0$), an expression for the excitation probability, $p_k \approx |a_1(\tau)|^2$ can be obtained \cite{Garanin:2002,Polkovnikov:2005,Grandi:2010}, where 
\be
a_1(\tau) \approx {-\frac{i}{2}} \int_0^\tau dt  \frac{\frac{dg}{dt} \sin k}{\sin^2 k +(g(t)-\cos k)^2} e^{i \varphi_k(t)}
\label{exctprob}
\ee 
and the dynamical phase, 
\be
\varphi_k(t) = 2\int_0^t dt' \epsilon_k(t') \label{dynph}
\ee
In the KZ scaling limits, $k \to 0, \tau \to \infty$ such that $k \tau^{\frac{\omega}{1+\omega}}$ is finite, we define 
\bea
X &=&  \frac{4 J (\tau-t_c) k^{\frac{1+\omega}{\omega}}}{\omega} ~,~ Y= \frac{t_c (g_i-1)^{-\frac{1}{\omega}}}{\tau-t_c} X 
\eea
For $X, Y \gg 1$, the  integral \eq{exctprob} is estimated using the method of stationary phase \cite{Bender:1978} in Appendix~\ref{app_spa}, and we find that, in general, $a_1(\tau)$ is a sum of terms that are exponential in $X$ (or $i X$ for even $\omega^{-1}$) and a term that decays algebraically in $X$.

\subsection{Symmetric quench}
\label{sec_symm}

\begin{figure}[t]
     \centering
     \begin{subfigure}{0.49\textwidth}
         \centering
         \includegraphics[width=1.0\textwidth]{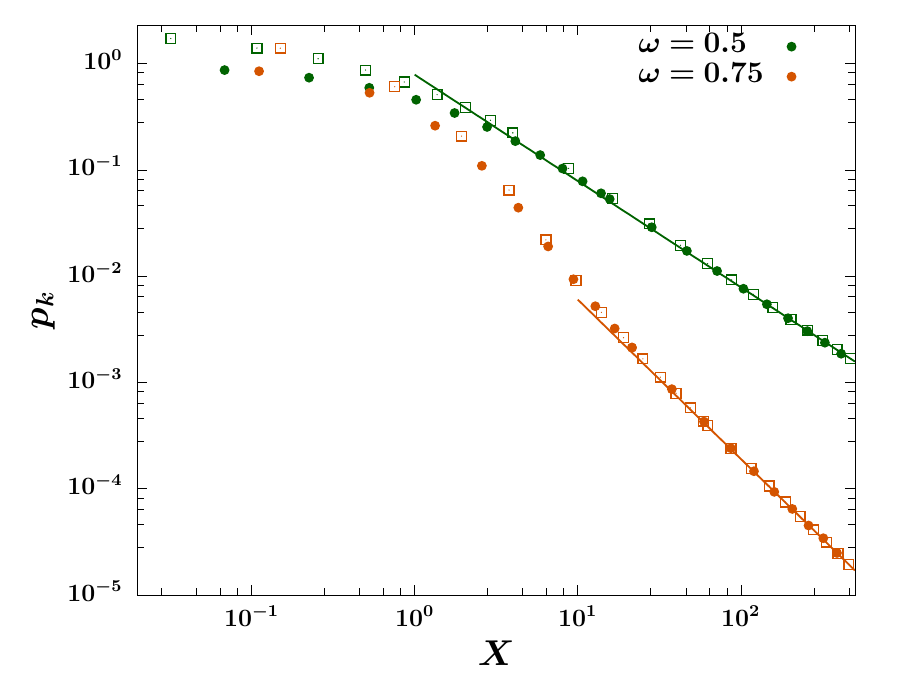}
         \caption{}
        \label{pk_b}
     \end{subfigure}
       \begin{subfigure}{0.49\textwidth}
         \centering
         \includegraphics[width=1.0\textwidth]{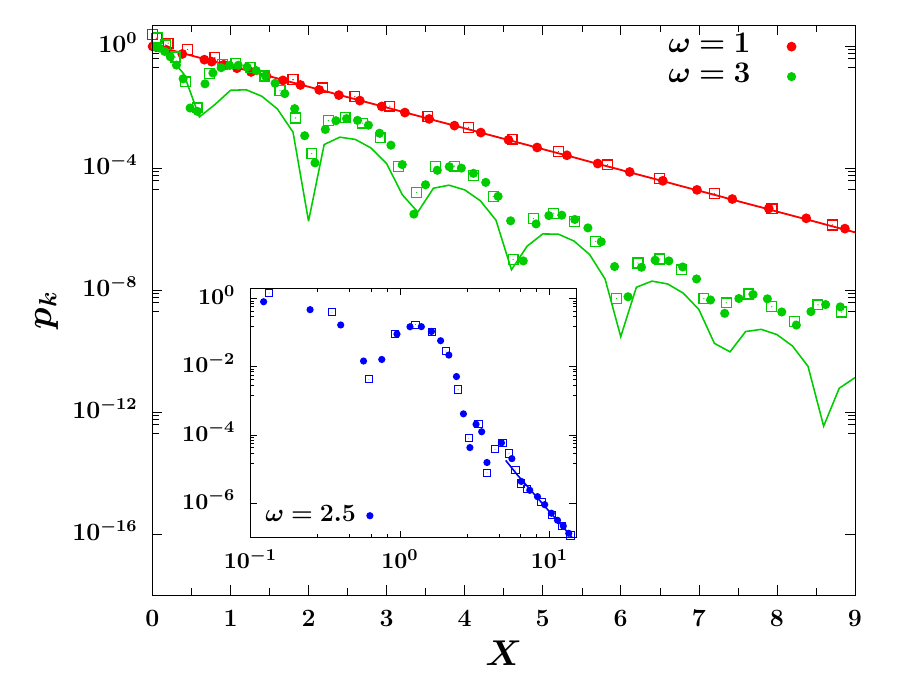}
         \caption{}
        \label{pk_a}
     \end{subfigure}
        \caption{Excitation probability, $p_k$ as a function of the scaled variable, $X=\frac{2 \tau k^{\frac{1+\omega}{\omega}}}{\omega}$ when the quench exponent (a) $\omega < 1$ and (b) $\omega \geq 1$. 
       The filled circles represent $p_k=|u'_k|^2$ which is obtained from the numerical solution of the differential equations (\ref{sm_ukpvkp}), the open squares correspond to $p_k \approx |a_1(\tau)|^2$, where $a_1(\tau)$ is calculated by solving the integral in Eq. \eq{sm_a1tau} numerically, and the solid lines denote the analytical expression \eq{pkeqt}. In both the  figures, $g_i=2$, $g_f=0$, $t_c=0.5\tau$ and $\tau=10^3$.}
\label{fig_apt}
\end{figure}

For $g_i=2$ and $t_c=0.5\tau$, the system spends equal time on either side of the critical point, and the general expression in Appendix~\ref{app_spa} for $a_1(\tau)$ simplifies to give
\be
p_k \approx \bigg|\frac{\pi}{3} e^{-X F(\omega) \sin(\frac{\pi}{2 \omega}) }  \cos\left(F(\omega) \cos \left(\frac{\pi}{2 \omega}\right) X\right)- \frac{ \Gamma (\omega +1)\cos(\frac{\pi \omega}{2})}{\omega^\omega} X^{-\omega} \bigg|^2 \label{pkeqt}
\ee
where, the scaled variable $X=Y= \frac{2 J \tau k^{\frac{1+\omega}{\omega}}}{\omega}$ is large, and 
\be
F(\omega)= \frac{1}{2}  \frac{\Gamma(\frac{3}{2}) \Gamma(\frac{1}{2 \omega})}{\Gamma(\frac{3}{2}+\frac{1}{2 \omega})} \label{fmain}
\ee 
The above expression is found to be in good agreement with the numerical evaluation of $|u'_k|^2$ using Eq. \eq{sm_ukpvkp} and the integral \eq{sm_a1tau}, as shown in Fig.~\ref{fig_apt}.


Equation \eq{pkeqt} shows that (i) when $\omega$ is an odd integer, $p_k$ is a periodic function of $X$ whose amplitude decays exponentially with $X$, except for linear quench where it is a purely exponential function, (ii) when $\omega=(2 n)^{-1}$ with $n=1, 2,...$, $p_k$ is a periodic function of $X$ which decays algebraically at large $X$, and (iii) for all other $\omega$, a crossover occurs from the damped cosine function to a power law decay. 
For $\omega \gg 1$, as $F(\omega) \sim \omega,  \frac{\Gamma (\omega +1)}{\omega^\omega} \sim \sqrt{2 \pi } e^{-\omega } \sqrt{\omega }$, the first term on the RHS of Eq. \eq{pkeqt} dominates the second term when 
$e^{-\frac{\pi X}{2}}  \gg e^{-\omega } \sqrt{\omega } X^{-\omega}$, or $X \ll \omega$. 
For $\omega \ll 1$, $F(\omega)\sim \sqrt{\frac{\pi }{2}} \omega^{3/2},\frac{\Gamma (\omega +1)}{\omega^\omega} \sim 1$ 
so that the first  term on the RHS of Eq. \eq{pkeqt} decays exponentially with a small rate until $X \ll \omega^{-3/2}$. Thus, in general, the excitation probability decays exponentially with $X$ (or $iX$) with crossover to a power law occurring when  $X$ is sufficiently large.

\subsection{Linear quench}

We now briefly discuss the effect of $g_i$ and $t_c$ for $\omega=1$; using the expressions in Appendix~\ref{app_spa}, we obtain
\bea
p_k &\approx& 
\frac{1}{4} \bigg|\frac{\pi}{3} \{e^{-\frac{\pi X}{4}}{+}e^{-X \frac{t_c}{(\tau-t_c)(g_i-1)} \frac{\pi}{4}}\} - \left\{1- \frac{(\tau-t_c) (g_i-1)}{t_c}  \right\} \frac{i}{X}\bigg|^2 \label{pklin}
\eea
where, $X=4 (\tau-t_c) k^2$.  {The above expression shows that if  $g_i -1 \neq \frac{t_c}{\tau-t_c}$, the algebraically decaying term in $X$ dominates at large $X$ and for such a quench, the excitation probability is not a Gaussian in momentum. Physically, this implies that if the control parameter changes rapidly in either phase, the energy in that phase changes rapidly and diabatic transition can occur with a larger probability. 
However, if the slope of the linear quench protocol is kept same above and below $t_c$, the power law term in Eq.  \eq{pklin} is identically zero and we recover the usual exponential law in $X$ \cite{Ivakhnenko:2023}. }

\section{Dephased kink-kink correlator} 
\label{sec_deph}

\begin{figure}[t]
     \centering
     \begin{subfigure}{0.48\textwidth}
         \centering
         \includegraphics[width=\textwidth]{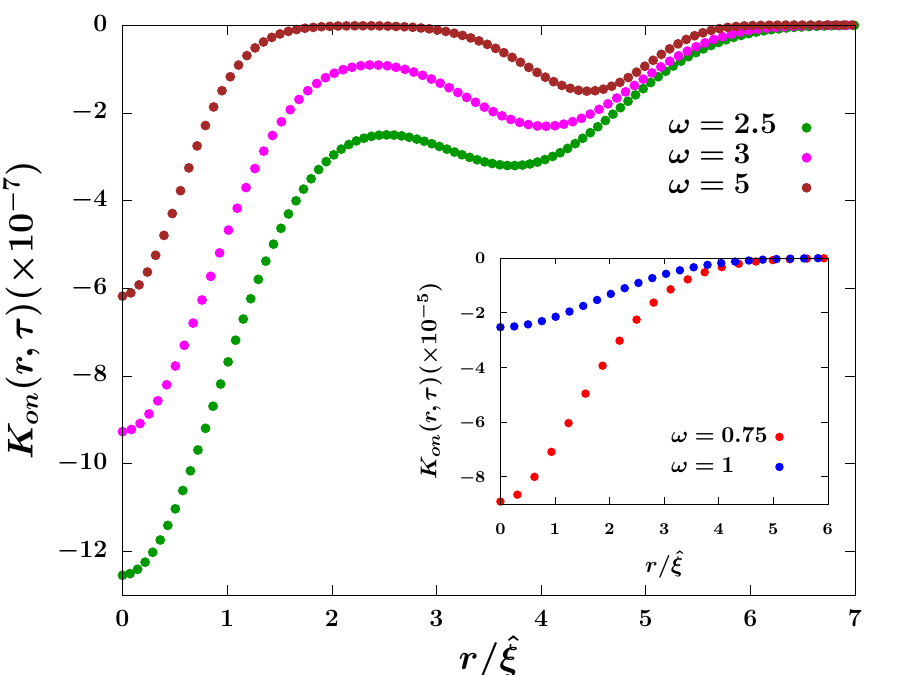}
         \caption{}
        \label{ukanaa}
     \end{subfigure}
     \begin{subfigure}{0.48\textwidth}
         \centering
         \includegraphics[width=\textwidth]{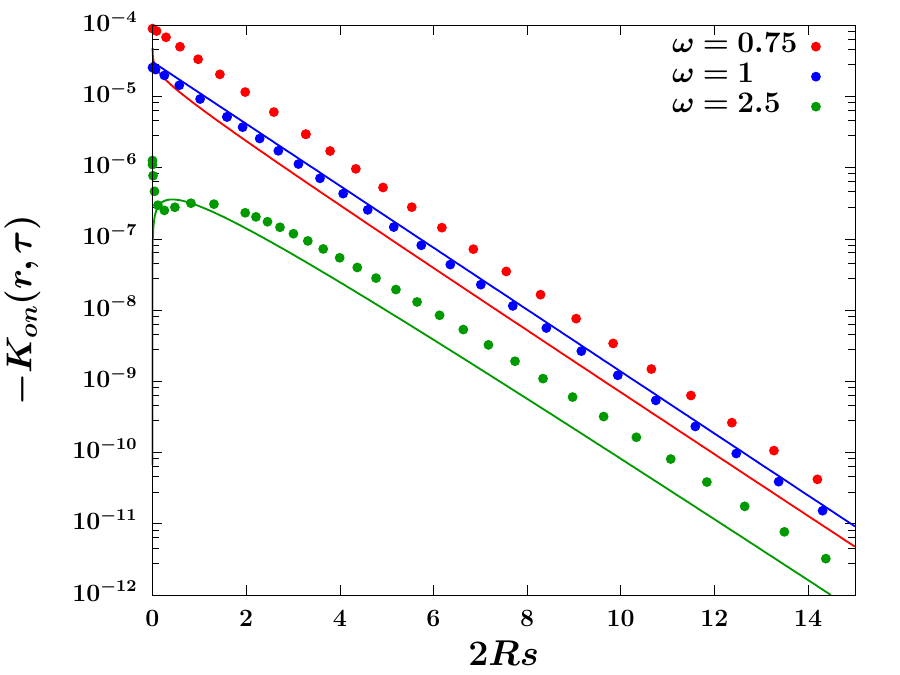}
        \caption{}
        \label{ukanac}
     \end{subfigure}
        \caption{(a) Dephased correlator, $K_{\text{on}}(r, \tau)$ as a function of kink-kink distance $r$ for $\omega>1$ (main) and $\omega \leq 1$ (inset). (b) Magnitude of dephased correlator at large distances shown as a function of the scaled variable, $2 R s$, where $R=r^{1+\omega}$, and $s$ defined by Eq.~\eq{app_defsx} is approximately $1.5 \times 10^{-3}, 1.5 \times 10^{-4}, 3 \times 10^{-10}$ for $\omega=0.75, 1, 2.5$, respectively, for the parameters in this figure. 
 For $\omega=2.5$, the large-distance behavior sets in at $r/{\hat \xi} \approx 6$ which is not visible in panel (a). 
 In both figures, the filled circles represent the dephased correlator \eq{kkon1} obtained numerically, while the lines in (b) show the analytical expression \eq{app_dep_ana}. In both the plots, $g_i=2, g_f=0, t_c=0.5\tau, \tau=10^3$, and ${\hat \xi}=\tau^{\frac{\omega}{1+\omega}}$.}
\label{ukana}
\end{figure}

As discussed in Sec.~\ref{qties}, the dephased correlator at the end of the quench is given by
\bea
K_{\text{on}}(r, \tau) &=& -   \left(\frac{1}{\pi} \int_0^\pi dk \cos(kr)\; |u'_k|^2 \right)^2 \label{kkon1}\\
&\approx& -   \left(\frac{1}{\pi} \int_0^\pi dk \cos(kr)\; p_k \right)^2  \label{kkon}
\eea
Figure~\ref{ukanaa} shows that the dephased correlator increases monotonically towards zero for $\omega \leq 1$, but it oscillates at short distances for superlinear quenches; here, we are interested in understanding how $K_{\text{on}}$ approaches zero. 

To find the large-$r$ behavior of the correlator, we need to consider the excitation probability $p_k$ at low momentum (or,  small $X$ for fixed quench time) which, as discussed in Sec.~\ref{sec_symm}, is exponential in $X$ (or $i X$) for all $\omega$. Using this approximation for $p_k$ in Eq. \eq{kkon}, the resulting integral is estimated in Appendix~\ref{app_deph} using the method of stationary phase. 
The result given by Eq. \eq{app_dep_ana} can be cast in the following scaling form, 
\bea
|K_{\text{on}}(r, \tau)| &\stackrel{r \gg 1}{\approx}& \frac{1}{{\hat \xi}^2} f_{\text{on}}\left(\frac{r}{{\hat \xi}} \right) \label{kksca}
\eea
where, the scaling function $f_{\text{on}}(\kappa) \sim \exp(-\kappa^{\omega+1})$. Figure~\ref{ukanac} shows that the numerical results for the correlation function agree quite well with Eq. \eq{app_dep_ana} at large distances for $\omega > 1$ 
and for almost the entire range of $r$ for $\omega \leq 1$.  Equation \eq{kksca} shows that the dephased correlator decays faster than an exponential for $\omega > 0$, and for linear quench, it decays as a Gaussian in $r$ \cite{Roychowdhury:2021}.

In the limit $\omega \to 0$, as shown in Fig.~\ref{prot_a}, $g(t) \approx g_i$ for $t < t_c$ and $g_f$ for $t > t_c$. But since the Hamiltonian \eq{hamil} at $g_f=0$ commutes with the longitudinal correlator, the initial correlation function does not evolve and remains close to that in the ground state at $g_i$. We thus expect that $K_{\text{on}}(r, \tau) \stackrel{\omega \to 0}{\to} K_{\text{eq}}(r)$, where the equilibrium correlator (given by Eq. \eq{kkon1} with $u_k'$ replaced by $u'_{k,eq}$)  decays exponentially with distance; the analytical result \eq{kksca} is indeed in agreement with this expectation (for a numerical verification, see Fig.~\ref{app_equq}). 

In the limit $\omega \to \infty$, as already mentioned in Sec.~\ref{sec_proto}, the protocol reduces to an instantaneous quench from $g_i$ to $g_c$. While longitudinal spin-spin correlation functions following an instantaneous quench have been studied \cite{Calabrese:2011,Calabrese:2012}, the corresponding results for transverse spin-spin correlation function which, due to duality can be related to longitudinal kink-kink correlator, do not appear to be known. However, Fig.~\ref{app_instq} suggests that the dephased correlator and total $K(r, \tau)$ decay exponentially with distance in this limit also.

\section{Off-diagonal kink-kink Correlator}
\label{offdia}

To understand the off-diagonal correlator defined in Sec.~\ref{qties}, we write  
\bea
K_{\text{off}}(r, \tau) &=&  \left| \frac{1}{\pi} \int_0^{\pi} dk \sin (kr) u_k'^{*}v_k' \right|^2 \label{koff2}\\
&=&  \left| \frac{1}{\pi}\textrm{Im} \left[\int_0^\pi dk \; e^{i k r}  u_k'^{*} v'_k \right] \right|^2 \\
&\approx&  \left| \frac{1}{\pi} \textrm{Im} \left[\int_0^\pi dk \; e^{i k r} \sqrt{p_k(1-p_k)}e^{i \Delta\varphi_k} \right] \right|^2 \label{Koff1}
\eea
where $|u'_k| \approx \sqrt{p_k}, |v'_k| \approx \sqrt{1-p_k}$, since the excitation probability does not evolve after the impulse regime (see, for e.g., Fig.~1a of \cite{Jindal:2025}), and $\Delta \varphi_k$ is the dynamical phase difference between modes that is accumulated during the impulse regime. In the KZ scaling limits, $k \to 0, \tau \to \infty, k \tau^{\frac{\omega}{1+\omega}}=k {\hat \xi}$ finite, 
 $\sqrt{p_k (1-p_k)} \approx \sqrt{p_k} \equiv {\cal P}(k {\hat \xi})$ where, as shown in Sec.~\ref{excit}, $\ln {\cal P}(y) \propto -y^{\frac{1+\omega}{\omega}}$.
 From Eq. \eq{dynph}, for low momentum modes, we have 
\bea
\Delta \varphi_k &=& \varphi_{k \to 0}-\varphi_0 \approx k \frac{\partial \varphi_k}{\partial k}=2 k \int_{{\hat t}}^{2 t_c-{\hat t}} dt' \frac{\partial \epsilon_k(t')}{\partial k} \\
&\approx& 4 J k^2 \int_{{\hat t}}^{2 t_c-{\hat t}} dt' \frac{g_c}{|g_c-g(t')|} 
\eea
Setting the prefactors that are independent of $k$ and $\tau$ to unity, we obtain the phase difference to be 
\bsn
{\Delta \varphi_k =}
k^2 \tau, & $\omega < 1$ \\
k^2 \tau \ln \tau, & $\omega=1$ \\
k^2 \tau^{\frac{2\omega}{1+\omega}}, & $\omega > 1$
\esn
which shows that on the KZ scale, the phase difference remains finite for superlinear quenches while it diverges with quench time for $\omega \leq 1$ \cite{Kolodrubetz:2012b}. 
Writing $\Delta \varphi_k= (k {\hat \xi}) (k \ell)$, we then obtain the dephasing length, 
\bsn
{\ell =\label{ellform}} 
\tau^{\frac{1}{1+\omega}}, & $\omega<1$ \label{ell1}\\
\sqrt{\tau}\ln \tau, & $\omega=1$ \\
\tau^{\frac{\omega}{1+\omega}}, & $\omega>1$
\esn
which is longer than KZ length for $\omega \leq 1$ and allows the system to eventually dephase, whereas the system does not dephase for superlinear quenches. 

\begin{figure}[t]
     \centering
     \begin{subfigure}{0.45\textwidth}
         \centering
         \includegraphics[width=\textwidth]{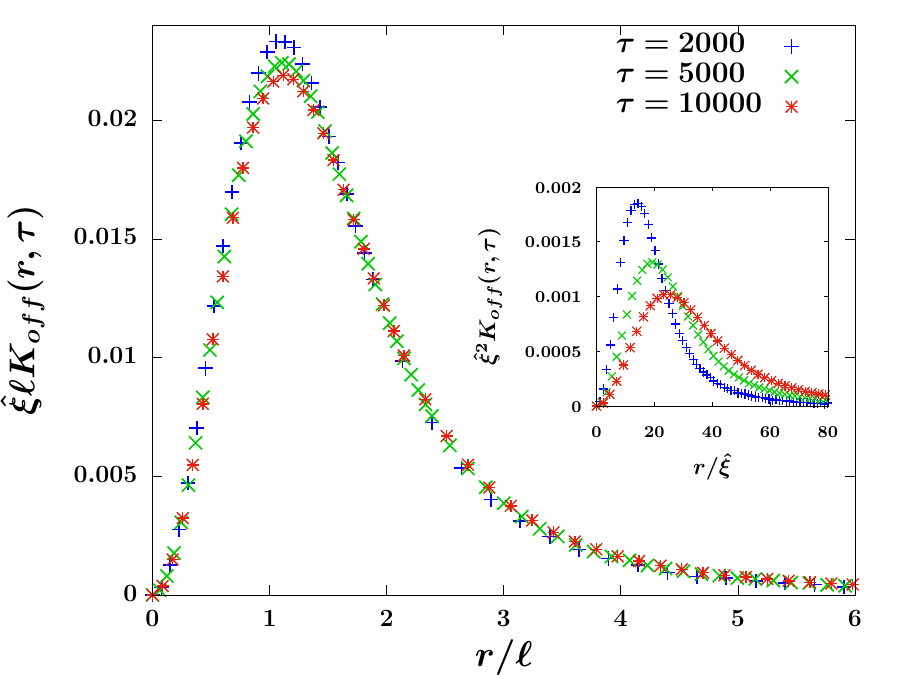}
         \caption{}
        \label{betaa}
     \end{subfigure}
     \begin{subfigure}{0.45\textwidth}
         \centering
         \includegraphics[width=\textwidth]{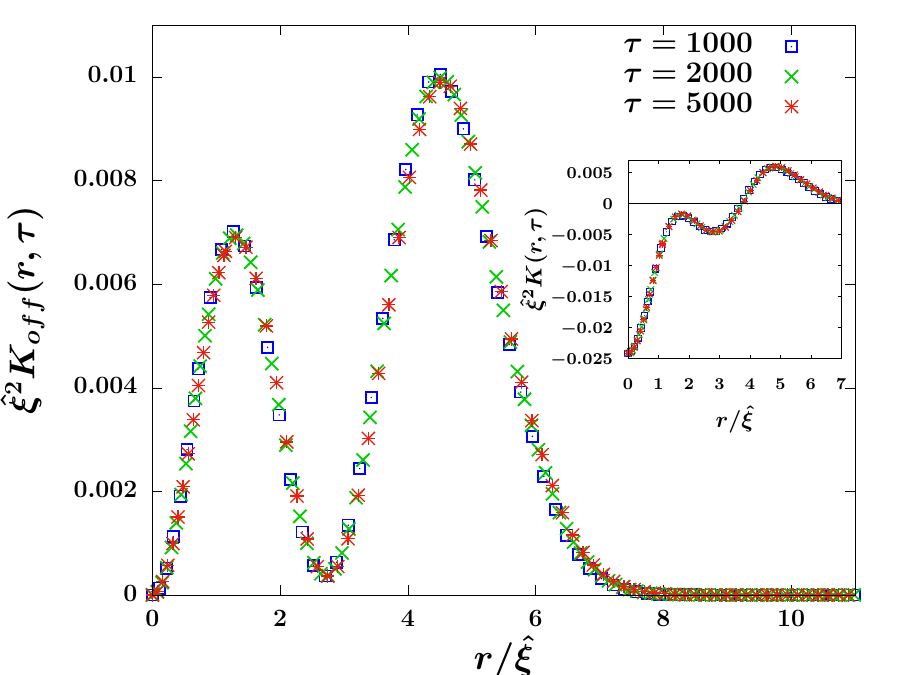}
        \caption{}
        \label{betac}
     \end{subfigure}
        \caption{Scaled off-diagonal kink-kink correlator as a function of scaled distance for (a) $\omega=0.5$ and (b) $\omega=2.5$ to support the scaling form \eq{Koffsc} when the dephasing length is given by Eq. \eq{ellform} (main panels). The inset in (a) shows that the off-diagonal correlator does not collapse with the KZ length scale for $\omega\leq 1$, while the inset in (b) shows that the total correlator depends only on the KZ scale for $\omega > 1$.  In both figures, $g_i=2, g_f=0, t_c=0.5\tau, {\hat \xi}=\tau^{\frac{\omega}{1+\omega}}$, and the data are obtained by numerically solving the differential equations (\ref{sm_ukpvkp}), and then integrating Eq. (\ref{koff2}).}
\label{beta}
\end{figure}

Using the above results in Eq. \eq{Koff1}, we get
\bea
K_{\text{off}}(r, \tau) &\approx&  \left|\frac{1}{\pi} \textrm{Im} \left[\int_0^\pi dk \; e^{i k r} {\cal P}(k  {\hat \xi}) e^{i k^2  {\hat \xi} \ell} \right] \right|^2 \label{Kofff}
\eea
Then, as shown in Appendix~\ref{app_offc}, the off-diagonal correlator can be written as
\bsn
{K_{\text{off}}(r,\tau)=\label{Koffsc}} 
\frac{1}{\hat{\xi} \ell} f_{\text{off}}^<\left(\frac{r}{\ell}\right) ~,~\omega \leq 1 \label{scafuncan1} \\
\frac{1}{\hat{\xi}^2} f_{\text{off}}^>\left(\frac{r}{\hat{\xi}}\right) ~,~\omega > 1 \label{scafuncan2}
\esn
The above scaling form is consistent with Eq.~(54) of \cite{Cincio:2007} for linear quench, and is tested in Fig.~\ref{beta} and Fig.~\ref{sm_offdiag} for various $\omega$. As shown in the inset of Fig.~\ref{betaa} for $\omega<1$, the data collapse fails if the KZ length is used instead of the dephasing length, while  an excellent data collapse is obtained with the dephasing length $\ell$ given in Eq. \eq{ell1}. 
In contrast, for $\omega>1$, the scaling collapse works well with the KZ length scale, as shown in Fig. \ref{betac}, as $\ell \sim {\hat \xi}$ in this regime. We also note that as for dephased correlator, the off-diagonal correlator has oscillations for superlinear quenches. 

\section{Optimal quench protocol}

\begin{figure}[t]
     \centering
     \begin{subfigure}{0.45\textwidth}
         \centering
         \includegraphics[width=\textwidth]{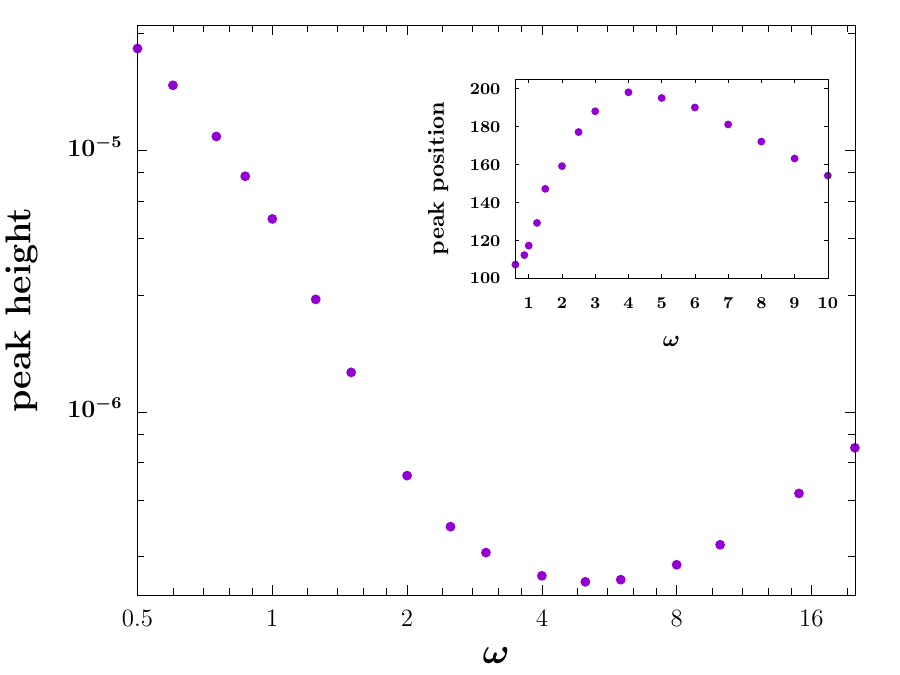}
         \caption{}
        \label{tw_a}
     \end{subfigure}
     \begin{subfigure}{0.45\textwidth}
         \centering
         \includegraphics[width=\textwidth]{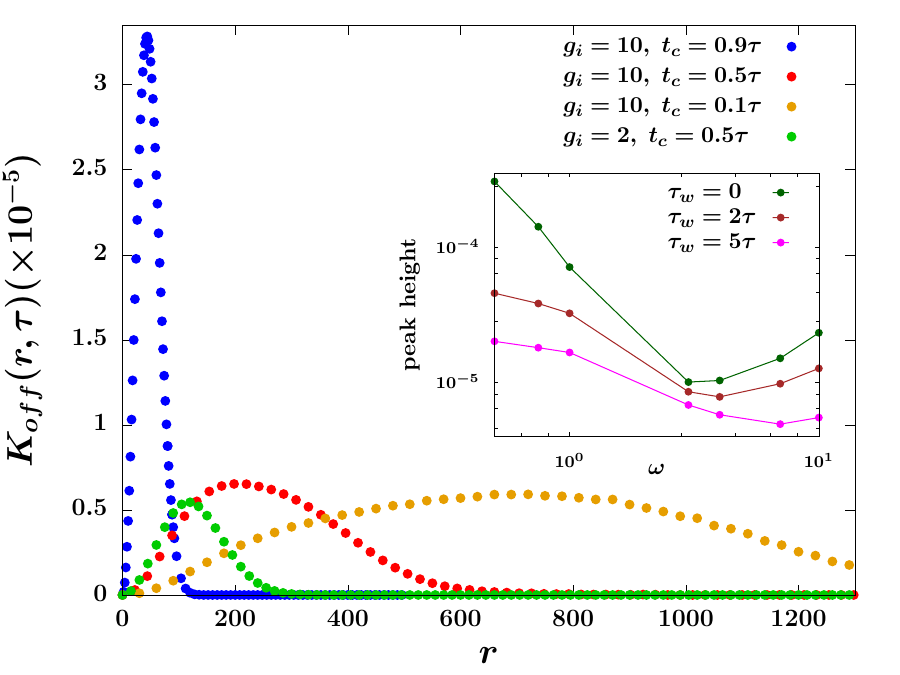}
         \caption{}
        \label{tw_b}
     \end{subfigure}
        \caption{(a) Peak height and the peak position in $K_{\text{off}}(r,\tau)$ as a function of $\omega$ for $g_i=2$, $\tau=10^3$, $t_c=0.5\tau$. (b) The main figure shows the off-diagonal correlator $K_{\text{off}}(r,\tau)$ for linear quench, and the inset shows the peak height as a function of quench exponent for various waiting times (see text for details) when $g_i=2, \tau=10^2, t_c=0.5\tau$; here, solid lines are guide to the eye.}
\label{tw}
\end{figure}

For a given $\omega$, as the longitudinal mean defect density, $\rho_d$ defined in Eq. \eq{rhoddef}, is proportional to ${\hat \xi}^{-1} \sim \tau^{-\frac{\omega}{1+\omega}}$, it decreases monotonically with quench time. For a fixed quench time $\tau$, one may expect the defect density to be a monotonically decreasing function of the quench exponent $\omega$ also: this is because when $\omega \to 0$, the kinks in the paramagnetic phase are passed on to the end of quench and when $\omega \to \infty$, the defects leftover after coarsening  at $g_c$ are observed at $t=\tau$ (refer to the discussion in Sec.~\ref{sec_deph} regarding these limits). However, there is an optimal exponent in the quench protocol \eq{nonlinprot} that minimizes the mean number of defects when the system is quenched from deep in the paramagnetic phase to the ferromagnetic phase \cite{Barankov:2008}; this conclusion holds for other $g_i, t_c$ also, as shown in Fig.~\ref{SM_defden}. 

Here, we ask if, for a given quench time, there is an optimal quench exponent for which the phase difference between modes is large enough that the phases at the end of the quench are essentially random and the system is dephased. To address this question, we first note that from Eq. \eq{Koffsc}, the peak of the off-diagonal correlator occurs at a typical distance $\ell$ and the peak height scales as $({\hat \xi} \ell)^{-1}$. We use the height of the peak in $K_{\text{off}}(r,\tau)$ as a proxy for the phase information as the off-diagonal correlator vanishes when the kink-kink correlator is dephased.  
 For a fixed $\tau$, Fig.~\ref{tw_a} shows that the height and the location of the first peak in  $K_{\text{off}}(r,\tau)$ vary  non-monotonically with the quench exponent $\omega$, and for the parameters in this figure, the peak height is minimum and the peak location is maximum when $\omega \sim 4-5$. We also note that although the peak height for $\omega<1$ is larger than that for $\omega \geq 1$, it falls faster with increasing $\omega$ for sublinear quenches than it grows for superlinear quenches, see main panel of Fig.~\ref{tw_a}.

We now enquire if, for a given $\omega$ and $\tau$, one could find parameters that could aid the randomization of the phases. Figure~\ref{tw_b} for linear quench suggests that to decrease the peak height in the off-diagonal correlator, one can either start close to the critical point (smaller $g_i$) or spend a larger fraction of the quench time in the ferromagnetic phase (smaller $t_c$). The latter observation is consistent with a protocol in which to enhance the dephasing length, the time-dependent transverse field is kept constant in the ferromagnetic phase at $g_w \in (0,1)$ for a waiting time $\tau_w$ \cite{Nowak:2021,Dziarmaga:2022}; this protocol and the resulting correlator are described in Appendix~\ref{app_halt}.
The inset of Fig. \ref{tw_b} shows the effect of $\tau_w$ on the peak height of the off-diagonal correlator, and we observe that when the peak height without halting is substantially large (as is the case for sublinear quenches), waiting in the ferromagnetic phase has a much stronger influence than when it is already small, and therefore, the halting time does not have a significant impact on the peak height for superlinear quenches and the optimal quench exponent is not substantially affected.

\section{Summary and open questions}

To understand the finite-time quench dynamics, here we considered a TFIM in one dimension with constant coupling $J$ and time-dependent transverse field $g$ that varies algebraically in the neighborhood of the critical point. 
Due to its solvability, much is known for linear quenches in the TFIM \cite{Dziarmaga:2010, Polkovnikov:2011, Dutta:2015, Campo:2014}, but it is not clear if and how these results are affected  for quenches that can not be linearized in the vicinity of the critical point. Unfortunately, the nonlinear quench problem does not appear to be exactly solvable, as discussed in Appendix~\ref{app_genl}; for this reason, our results are obtained using an adiabatic perturbation theory and an analytic argument that are verified against exact numerical solution of Eq. \eq{sm_ukpvkp}. 

Our results in Sec.~\ref{excit}  for the excitation probability following a nonlinear quench are essentially the same as in \cite{Garanin:2002}, but here we have also discussed the effect of initial transverse field and the time spent in a phase, and  more importantly, we have used these results to find an analytical expression for the dephased correlator.  
We find that for nonlinear quenches, the dephased correlator always decays super-exponentially which adds to the discussion in \cite{Roychowdhury:2021} that the AIA approximation \cite{Damski:2005,Damski:2006} does not capture the behavior of the correlation functions. More precisely, our analysis shows that at large distances, the dephased correlator decays as a compressed exponential with a continuously varying exponent - this is a rather unusual result as one may have expected that the nonuniversal aspects of the protocol will get absorbed in the definition of the KZ scale, resulting in a universal scaling function. For the off-diagonal correlator, we find that on the KZ length scale, {the system eventually dephases for sublinear and linear quenches, and this process can be aided if the transverse field is kept constant for some time in the ferromagnetic phase. But the system fails to dephase for superlinear quenches. Interestingly, for a fixed quench time, the optimal quench exponent for the mean defect density and measures of dephasing are found to lie in the same range.}
 
The present work can be extended in several directions: following a linear quench, while the transverse spin-spin correlation function decays as a Gaussian, the longitudinal spin-spin correlation function decays exponentially \cite{Cincio:2007}, and it might be interesting to understand if the latter behavior is affected by nonlinear quenches. At the end of linear quench, if the system is freely evolved, at late times, the state of TFIM  
 can be described by generalized Gibbs ensemble \cite{Rigol:2007,Kolodrubetz:2012b,Vidmar:2016}, and one may enquire if the long time state or approach to it carries any signature of the nonlinearity in the quench protocol; in particular, as the system does not dephase for superlinear quenches, it would be interesting to investigate the long time behavior in this case. We also mention that for linear quench, the entanglement entropy of a subsystem increases with its size saturating to  a value proportional to the logarithm of the KZ length scale but the crossover occurs at the dephasing length scale \cite{Cincio:2007}; we expect that analogous results hold for the nonlinear quench protocol considered here, but this expectation remains to be tested.  The dynamical quantum phase transition \cite{Heyl:2013,Heyl:2018} which has been studied after a sudden and linear quench  could also be explored for nonlinear quenches. \\

\clearpage
\appendix
\makeatletter
\renewcommand{\@seccntformat}[1]{Appendix \csname the#1\endcsname\quad}
\makeatother
\renewcommand{\thesection}{\Alph{section}}
\numberwithin{equation}{section}

\section{Excitation probability}
\label{app_spa}

To find the excitation probability $p_k$, 
we now work in the KZ scaling regime where $k \to 0, \tau \to \infty$ such that $k \tau^{\frac{\omega}{1+\omega}}$ is finite. 
From Eq. \eq{exctprob}, we can write 
\bea
a_1(\tau)&\stackrel{k \to 0}{\approx}& 
\frac{i k}{2} \int_{-1}^{0}   \frac{d \lambda}{k^2 +\lambda^2} e^{ -\frac{i  4 J (\tau-t_c)}{\omega} \int_0^\lambda \frac{d \Lambda}{ (-\Lambda)^{\frac{\omega-1}{\omega}}} \sqrt{k^2+\Lambda^2}} 
+ \frac{i k}{2} \int_{0}^{g_i-1}  \frac{d \lambda }{k^2 +\lambda^2} e^{- \frac{i  4 J (g_i-1)^{-\frac{1}{\omega}}t_c}{\omega} \int_0^\lambda \frac{d \Lambda}{ \Lambda^{\frac{\omega-1}{\omega}}} \sqrt{k^2+\Lambda^2}} \label{app_apt1}\\
&\approx& \frac{i}{2} \int_{0}^{\infty}  \frac{dp}{1 +p^2} e^{i X  f(p)} +\frac{i}{2} \int_{0}^{\infty}  \frac{dp}{1 +p^2} e^{-i Y  f(p) } \label{sm_a1tau}
\eea
where, the last expression is obtained in the aforementioned scaling limits with 
\bea
X &=&  \frac{4 J (\tau-t_c) k^{\frac{1+\omega}{\omega}}}{\omega} ~,~ Y=\frac{t_c (g_i-1)^{-\frac{1}{\omega}}}{\tau-t_c} X
\eea
and 
\bea
f(p) &=& \int_0^p dP P^{\frac{1-\omega}{\omega}} \sqrt{1+P^2} 
\eea
For $X \gg 1$, the integrals in Eq. \eq{sm_a1tau}  can be estimated using a stationary phase approximation where the contribution to these integrals comes from the saddle points, singularities of the integrand and the end points \cite{Bender:1978}. 

Consider the integral, $I(X)=\int_{0}^{\infty}  \frac{dp}{1 +p^2} e^{i X  f(p)}$: we first note that in the 
complex $p$-plane, $f'(p)$ vanishes at the saddle points given by $p^*=\pm i, 0$ for $\omega  < 1$ and $p^*=\pm i$ for  $\omega \geq 1$; the function $f(p)$ is also non-analytic with branch points at $\pm i$ for all $\omega$, and at zero except when $\omega^{-1}$ is an integer (since $f(p) \stackrel{p \to 0}{\sim} p^{\frac{1}{\omega}}$). The integrand of $I$ also has simple poles at $\pm i$, and the lower limit of the integral is finite. Below we consider the contribution of these singularities to the integral $I$. 

We first expand $f(p)$ about the saddle point $+i$ by writing $p=i+z, z=\rho e^{i \theta}$, and obtain 
\bea
f(p) \approx f(i) + \frac{2}{3} \sqrt{2} \rho^{3/2} e^{i [\frac{\pi}{2} (\frac{1-\omega}{\omega}+\frac{1}{2})+\frac{3}{2} \theta]} =f(i) -\frac{2}{3} \sqrt{2} \rho^{3/2}
\eea
on choosing $\theta$ such that the real part of the second term is minimized for steepest descent. Changing the variables from $p$ to $\rho$, we then obtain 
\bea
I_{+}(X)&\approx&  \frac{e^{i X f(i)}}{2 i} \int_{0}^\infty \frac{d\rho}{\rho} e^{-\frac{2 \sqrt{2}  X}{3} \rho^{3/2}} = -\frac{\pi}{6} e^{i X f(i)} 
\eea 
where, 
\bea
f(\pm i) &=&  \frac{e^{\pm \frac{i \pi}{2 \omega}}}{2}  \frac{\Gamma(\frac{3}{2}) \Gamma(\frac{1}{2 \omega})}{\Gamma(\frac{3}{2}+\frac{1}{2 \omega})} \label{phiatSP}
\eea
on choosing a contour around the pole at $\rho=0$ and the branch cut at $(-\infty,0]$ \cite{Garanin:2002}. 
Similarly, the saddle point at $-i$ gives 
\bea
I_{-}(X) \approx -\frac{\pi}{6} e^{i X f(-i)}
\eea
on noting that the contour for the singularity at $-i$ is closed in the direction opposite to that for $+i$. The terms contributing  to Eq. \eq{sm_a1tau} due to the saddle point at $\pm i$ are given by the integrals that converge with increasing $X$. We therefore have
\bsn
{a_1^{saddle} =} 
\frac{i}{2} [I_+(X) + I_-(-\frac{t_c (g_i-1)^{-\frac{1}{\omega}} }{\tau-t_c} X)] , \sin(\frac{\pi}{2 \omega}) > 0 \\
\frac{i}{2} [I_-(X) + I_+(- \frac{t_c (g_i-1)^{-\frac{1}{\omega}} }{\tau-t_c} X)] ,  \sin(\frac{\pi}{2 \omega}) < 0 \\
\frac{i}{2} [I_+(X)+I_-(X)+I_-(- \frac{t_c (g_i-1)^{-\frac{1}{\omega}} }{\tau-t_c} X)+ I_+(- \frac{t_c (g_i-1)^{-\frac{1}{\omega}} }{\tau-t_c} X)] , \sin(\frac{\pi}{2 \omega}) = 0
\esn
which simplifies to yield
\bsn
{a_1^{saddle} \sim \label{app_saddlec}} 
 -\frac{i}{2} \frac{\pi}{3}[ e^{i X e^{\frac{i \pi}{2 \omega}} F(\omega)}+ e^{-i X e^{-\frac{i \pi}{2 \omega}} G(\omega)}] , \sin(\frac{\pi}{2 \omega}) > 0 \label{app_saddlecp}\\
-\frac{i}{2} \frac{\pi}{3} [ e^{i X e^{-\frac{i \pi}{2 \omega}} F(\omega)}+e^{-i X e^{\frac{i \pi}{2 \omega}} G(\omega)} ] , \sin(\frac{\pi}{2 \omega}) < 0 \\
-\frac{i}{2} \frac{\pi}{3}[2 e^{i X (-1)^n F(\omega)}+2 e^{-i X (-1)^n G(\omega)}] ,  \sin(\frac{\pi}{2 \omega}) = 0
\esn
where, in the last expression, $\omega=(2 n)^{-1}, n=1, 2, ...$ and 
\bea
F(\omega) &=&   \frac{\Gamma(\frac{3}{2}) \Gamma(\frac{1}{2 \omega})}{2 \Gamma(\frac{3}{2}+\frac{1}{2 \omega})} \label{app_Fdef}\\
G(\omega) &=&\frac{t_c (g_i-1)^{-\frac{1}{\omega}} }{\tau-t_c} F(\omega) \label{app_Gdef}
\eea

We next consider the contribution from the origin. On writing $p=\rho e^{i \theta}$ and expanding $f(p)$ about the origin, we obtain 
\bea
I_0(X) &\approx& e^{\frac{i \pi \omega}{2}} \int_{0}^\infty d\rho e^{- \omega X \rho^{\frac{1}{\omega}}} \\
&=& e^{\frac{i \pi \omega}{2}} (\omega  X)^{-\omega } \Gamma (\omega +1)
\eea
where, $\sin (\frac{\theta}{\omega})=1$ for steepest descent for $X > 0$. The total contribution to $a_1(\tau)$ due to the origin is then given by
\bea
a_1^{origin} &\sim& \frac{i}{2} [I(X)+I(-X \frac{t_c (g_i-1)^{-\frac{1}{\omega}} }{\tau-t_c})] \\
&=& \frac{i}{2} e^{\frac{i \pi \omega}{2}} (\omega)^{-\omega } \Gamma (\omega +1) X^{-\omega} \left[1+ \left(\frac{t_c}{\tau-t_c} \right)^{-\omega}(g_i-1)e^{-i \pi \omega} \right] \label{app_endptc}
\eea
Adding the contributions given by Eqs. \eq{app_saddlec} and \eq{app_endptc}, we obtain the total $a_1(\tau)$.

\section{Dephased correlator}
\label{app_deph}

To find the dephased correlator \eq{kkon}, we consider the Fourier transform of the excitation probability, ${\tilde p}(r)=\int_0^\pi dk e^{i k r} p_k$. Then for large $r$, as  small $k$ behavior of the excitation probability is required, we can ignore the algebraically decaying terms given by Eq. \eq{app_endptc}.  
When $\sin(\frac{\pi}{2 \omega}) > 0$, Eq. \eq{app_saddlecp} then gives the excitation probability to be
\bea
p_k &\approx& \frac{\pi^2}{36} [e^{-2 X \sin(\frac{\pi}{2 \omega}) F(\omega)}+ e^{-2 X \sin(\frac{\pi}{2 \omega}) G(\omega)}
+ e^{- X (F+G) \sin(\frac{\pi}{2 \omega})} e^{i X (F+G) \cos(\frac{\pi}{2 \omega})} +h.c.] \label{app_pkcorr}
\eea
where, $F$ and $G$ are, respectively, given by Eqs.  \eq{app_Fdef} and \eq{app_Gdef}. Thus, to find ${\tilde p}(r)$, we need to evaluate  integrals of the following form:
\bea
C_j &=& \frac{\pi^2}{36} \int_0^\pi dk e^{i k r} \; e^{-X b_j} \\
&=& \frac{\pi^2}{36}\frac{c_\tau \omega}{1+\omega} \int_0^{\infty}  \frac{dX}{X^{\frac{1 }{\omega +1}}} e^{i c_\tau X^{\frac{\omega }{\omega +1}} r} \; e^{-X b_j}  \\
&=& \frac{\pi^2}{36} \frac{c_\tau \omega r^\omega}{1+\omega}  \int_0^{\infty}  \frac{dx}{x^{\frac{1 }{\omega +1}}} e^{R [i c_\tau x^{\frac{\omega }{\omega +1}} - b_j x]} \label{app_EIdef}
\eea
where, $c_\tau=(\frac{\omega}{4 J (\tau-t_c)})^{\frac{\omega}{1+\omega}} \propto {\hat \xi}^{-1}, R=r^{\omega+1}$, and $b_j$ is, in general, complex.

For large $R$, we can find an approximate expression for $C_j$ using a stationary phase approximation. We first note that in the complex $x$-plane, the integrand of $C_j$ has branch point or poles at the origin, and the contribution from the origin $C_j^{origin} \propto \int_0 dx \; x^{-\frac{1 }{\omega +1}}$ which vanishes as $\omega > 0$. Writing the exponential factor in the integrand of $C_j$ as $e^{R h_j(x)}$, where $h_j(x)=i c_\tau x^{\frac{\omega }{\omega +1}} -b_j x$, 
we find that there are saddle points given by 
\bea
x_j^*= \left(\frac{i c_\tau \omega }{(\omega +1) b_j} \right)^{\omega +1} \label{app_saddles}
\eea
Then expanding $h_j(x)$ about the saddle point to quadratic order in $x-x_j^*=\rho e^{i \theta_j}$, and 
on carrying out the Gaussian integral over $\rho$, we find that the contribution from the saddle point(s) is given by
\bea
C_j^{saddle}&\sim& \sum_{x_j^*} \frac{\pi^2}{36} \frac{c_\tau \omega r^\omega}{1+\omega} \frac{e^{R h_j(x_j^*)}}{x_j^{*\frac{1 }{\omega +1}}}  e^{i \theta_j} \sqrt{\frac{2 \pi}{R |h_j''(x_j^*)|}} \label{app_Csgenl}
\eea
on choosing $\theta_j$ such that $ e^{i 2 \theta_j} h_j''(x_j^*)=- |h_j''(x_j^*)|$, and where,
\bea
h_j(x_j^*) &=& \frac{i c_\tau}{\omega+1} \left(\frac{i c_\tau \omega}{(\omega+1) b_j} \right)^\omega \label{gxs} \\
h_j''(x_j^*) &=& -\frac{i c_\tau \omega }{(\omega +1)^2} (x_j^*)^{-\frac{\omega+2}{\omega +1}} \label{g2xs}
\eea

We now consider each of the terms on the RHS of Eq. \eq{app_pkcorr}. For the first term, $b_1=2 \sin(\frac{\pi}{2 \omega}) F$; since $b_1$ is real, from Eq. \eq{app_saddles}, we find that the saddle points occur at $x_1^*  \propto i^{\omega+1} =e^{i \Theta_1}$ where, $\Theta_1=(2 m+1) (\omega+1) \frac{\pi}{2}$ 
 with $0 < \Theta_1 <  2\pi$ and $m=0, 1, 2...$. Then
 \bea
 h_1(x_1^*) 
&=& \frac{e^{i \Theta_1} c_\tau}{\omega+1} \left(\frac{ c_\tau \omega }{(\omega +1) 2 \sin(\frac{\pi}{2 \omega}) F} \right)^\omega
 \eea
Assuming that the contour can be deformed such that it passes through the saddle points where the integral $C_1$ does not diverge with $R$, the function $h_1$ is evaluated at those saddle points where $\cos \Theta_1 < 0$.  Similarly,  the second term in Eq. \eq{app_pkcorr} gives $b_2=2 \sin(\frac{\pi}{2 \omega}) G$ which yields 
\bea
 h_2(x_2^*) 
&=& \frac{e^{i \Theta_1} c_\tau}{\omega+1} \left(\frac{ c_\tau \omega }{(\omega +1) 2 \sin(\frac{\pi}{2 \omega}) G} \right)^\omega
 \eea
For the third term in Eq. \eq{app_pkcorr}, $b_3=-i e^{\frac{i \pi}{2 \omega}} (F+G)$ so that the saddles occur at $x_3^* \propto e^{i (2m+1) \pi (\omega+1)} e^{-\frac{i \pi (\omega+1)}{2 \omega}}$ and Eq. \eq{gxs} then gives
 \bea
 h_3(x_3^*) 
 &=&   \frac{e^{i \Theta_3} c_\tau}{\omega+1} \left(\frac{c_\tau \omega }{(\omega +1) (F+G)} \right)^\omega
 \eea
 where $\Theta_3=(2m+1) \pi \omega$ with non-negative integer $m$ chosen  such that $\cos \Theta_3 < 0$. Finally, for the fourth term in Eq. \eq{app_pkcorr}, $b_4=i e^{-\frac{i \pi}{2 \omega}} (F+G)$,  the saddles occur at $x_4^* \propto e^{\frac{i \pi (\omega+1)}{2 \omega}}$, and Eq. \eq{gxs} yields
 \bea
 h_4(x_4^*) &=&  -\frac{c_\tau}{\omega+1} \left(\frac{c_\tau \omega }{(\omega +1) (F+G)} \right)^\omega 
 \eea

 When $\sin(\frac{\pi}{2 \omega}) < 0$, the excitation probability is obtained on replacing $\sin(\frac{\pi}{2 \omega})$ in Eq. \eq{app_pkcorr} by its modulus, and the rest of the discussion above remains unchanged. 
For $\sin(\frac{\pi}{2 \omega})=0$, $p_k \approx \frac{\pi^2}{36} [2+e^{i X (-1)^n (F+G)}+h.c.]$ for $\omega=(2n)^{-1},  n=1,2,3...$. The first term in $p_k$ gives $\int_0^\pi dk \; 2 \cos(k r) =\frac{2 \sin(\pi r)}{r}$. For odd $n$, $h(x_j^*) \propto i (i/b)^\omega$ is $i$ for $b=-i (-1)^n (F+G)$ and $i (-1)^\omega=i \cos(\pi \omega)-\sin (\pi \omega)$ for $b=i (-1)^n (F+G)$. For even $n$, these two behavior are interchanged. Thus, for either $n$,  we obtain
 \bea
h(x_j^*)&=& \frac{c_\tau e^{i \Theta}}{\omega+1} \left(\frac{ c_\tau \omega }{(\omega +1) (F+G) } \right)^\omega
\eea
where $\Theta=(2m+1)\pi w +\pi/2$ with $m=0,1,2,...$.

Thus, we find that at large kink-kink distance, the magnitude of the dephased correlator \eq{kkon} is given by 
\bea
|K_{\text{on}}(r)| &\stackrel{r \gg 1}{\approx}&  \left(\sum_{j=1}^4 \textrm{Re}\left[\frac{1}{36} \frac{c_\tau \omega r^\omega}{1+\omega} \frac{e^{R h_j(x_j^*)}}{x_j^{*\frac{1 }{\omega +1}}}  e^{i \theta_j} \sqrt{\frac{2 \pi}{R |h_j''(x_j^*)|}} \right] \right)^2 \label{app_dep_ana1}
\eea
Furthermore, as the behavior of the correlator is dominated by the slowest decaying term in Eq. \eq{app_dep_ana1}, we can write 
\bea
|K_{\text{on}}(r)| &\approx& \left(\textrm{Re}\left[\frac{1}{36} \frac{c_\tau \omega r^\omega}{1+\omega} \frac{e^{R s(x^*)}}{x^{*\frac{1 }{\omega +1}}}  e^{i \theta} \sqrt{\frac{2 \pi}{R |s''(x^*)|}} \right] \right)^2
\label{app_dep_ana}
\eea
where, 
\be
s(x^*)=\min_j h_j(x_j^*) \label{app_defsx}
\ee 
For the parameters in Fig.~\ref{ukanac}, all the four $h_j$'s are equal for $\omega=1$, while for $\omega \neq 1$,  $h_4(x^*)$ is the smallest. From Eqs. \eq{gxs} and \eq{g2xs}, as $h_j(x_j^*) \propto c^{1+\omega}_\tau \propto {\hat \xi}^{-1-\omega}$ and $|h_j''(x_j^*)| \propto c^{-1-\omega}_\tau$, from Eq. \eq{app_dep_ana}, it follows  that the dephased correlator is of the scaling form in Eq. \eq{kksca}.    


\section{Dephased correlator in limiting cases} 

\begin{figure}[h!]
     \centering
     \begin{subfigure}{0.48\textwidth}
         \centering
         \includegraphics[width=\textwidth]{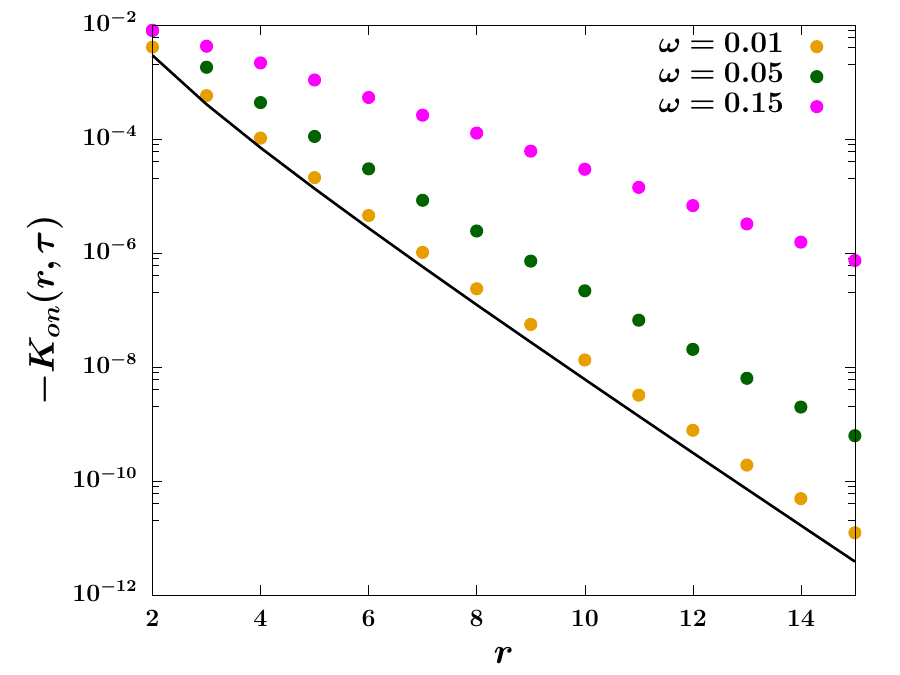}
         \caption{}
        \label{app_equq}
     \end{subfigure}
     \begin{subfigure}{0.48\textwidth}
         \centering
         \includegraphics[width=\textwidth]{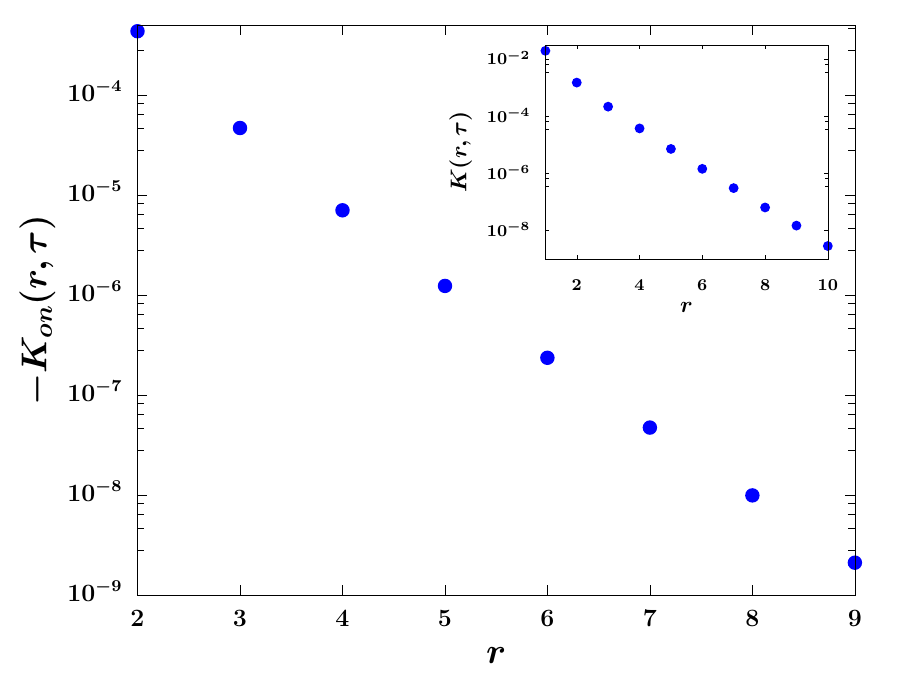}
         \caption{}
        \label{app_instq}
     \end{subfigure}
  \caption{(a) Magnitude of the dephased correlator at the end of the quench in the limit $\omega \to 0$ to show that it approaches the kink-kink correlator in the ground state at $g_i=2$ (black line) which is obtained on numerically integrating Eq. \eq{kkon1} after replacing $u_k'$ by $u'_{k,eq}$, and decays exponentially with distance $r$. 
(b) Magnitude of the dephased correlator (main) and the total correlator (inset) after an instantaneous quench from $g_i$ to $g_c$ followed by free evolution for a time $\tau$. 
  The data are obtained by numerically solving the differential equations (\ref{sm_ukpvkp}) and then integrating Eq. (\ref{kkon1}). In both figures, $g_i=2$, $g_f=0$, $\tau=10^3$ and $t_c=0.5\tau$ in the quench protocol (\ref{nonlinprot}).}
\label{SM_om0}
\end{figure}

%
\section{Off-diagonal correlator}
\label{app_offc}

To show that the off-diagonal correlator \eq{Kofff} has the scaling form \eq{Koffsc}, consider the integral $\int_0^\pi dk \; e^{i k r}  e^{i k^2  {\hat \xi} \ell} \sqrt{p_k}$ where $p_k$ is given by Eq. \eq{app_pkcorr}, and one can work out the integral in a manner similar to that in Appendix~\ref{app_deph}. Here, we merely sketch the calculation leading to Eq. \eq{Koffsc}, and for this purpose, we consider 
\bea
D &=& \int_0^\pi dk \; e^{i k r}  e^{i k^2  {\hat \xi} \ell} e^{-A (k {\hat \xi})^{\frac{1+\omega}{\omega}}}
\eea
where, the factors constant in $k, {\hat \xi}$ are subsumed in $A$ which, in general, is complex. Changing the dummy variable $k$ to $Q=k {\hat \xi}+L$ where $L=\frac{r}{2 \ell}$, we obtain
\bea
D &=&  \frac{e^{-\frac{i r^2}{4 {\hat \xi} \ell}}}{{\hat \xi}} \int_{0}^\infty dQ e^{\frac{i \ell (Q+L)^2}{{\hat \xi}} -A Q^{\frac{1+\omega}{\omega}}} \\
&=& \frac{\sqrt{\pi}}{\sqrt{{\hat \xi} \ell}} \times \sqrt{\frac{\ell}{{\hat \xi}}}  e^{-\frac{i r^2}{4 {\hat \xi} \ell}} e^{i \theta} e^{H(Q_*)} \sqrt{\frac{2}{|H''(Q_*)|}} \label{app_Dsad}
\eea
where, 
\bea
H &=& \frac{i \ell (Q+L)^2}{{\hat \xi}} -A Q^{\frac{1+\omega}{\omega}} \\
H' &=& \frac{2 i \ell (Q+L)}{{\hat \xi}} - \frac{A (1+\omega)}{\omega} Q^{\frac{1}{\omega}} \label{Hp}\\
H'' &=& \frac{2 i \ell }{{\hat \xi}} - \frac{A (1+\omega)}{\omega^2} Q^{\frac{1-\omega}{\omega}} \label{Hp2}
\eea
and $e^{i 2 \theta}H''(Q_*)=-|H''(Q_*)|$. 

For $\omega \leq 1$, as $\ell \gg \xi$, neglecting the second term in Eq. \eq{Hp}, we get the saddle point $Q_* \approx -L, H(Q_*) \approx -A (-L)^{\frac{1+\omega}{\omega}}, H''(Q_*) \approx \frac{2 i \ell }{{\hat \xi}}$. For $\omega > 1$, $\ell \sim \xi$, but one can still neglect the second term in Eq. \eq{Hp} as $\omega > 1$ and we expect $Q_*$ to be large; thus, we obtain the same expressions for $Q_*, H(Q_*), H''(Q_*)$ as above for $\omega \leq 1$. Using these results  in Eq. \eq{app_Dsad}, we immediately find that 
\be
D \approx  \frac{1}{\sqrt{{\hat \xi} \ell}} {\cal D} \left( \frac{r}{\ell} \right)
\ee
and from which Eq. \eq{Koffsc} follows.

\section{Scaling collapse for off-diagonal correlator}

\begin{figure}[ht]
     \centering
     \begin{subfigure}{0.4\textwidth}
         \centering
         \includegraphics[width=\textwidth]{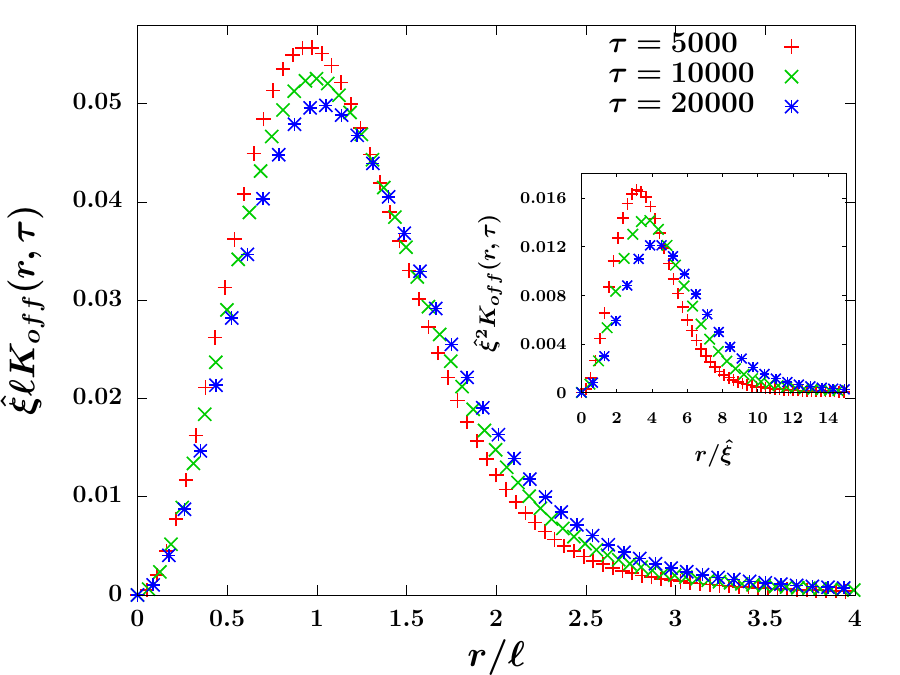}
         \caption{}
     \end{subfigure}
     \begin{subfigure}{0.4\textwidth}
         \centering
         \includegraphics[width=\textwidth]{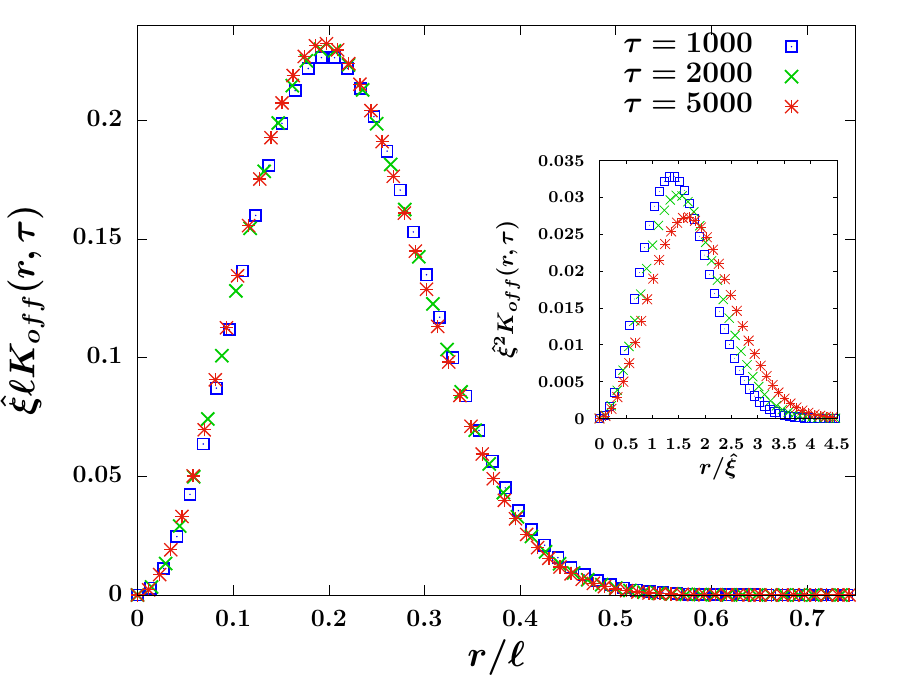}
         \caption{}
     \end{subfigure}
     \begin{subfigure}{0.4\textwidth}
         \centering
         \includegraphics[width=\textwidth]{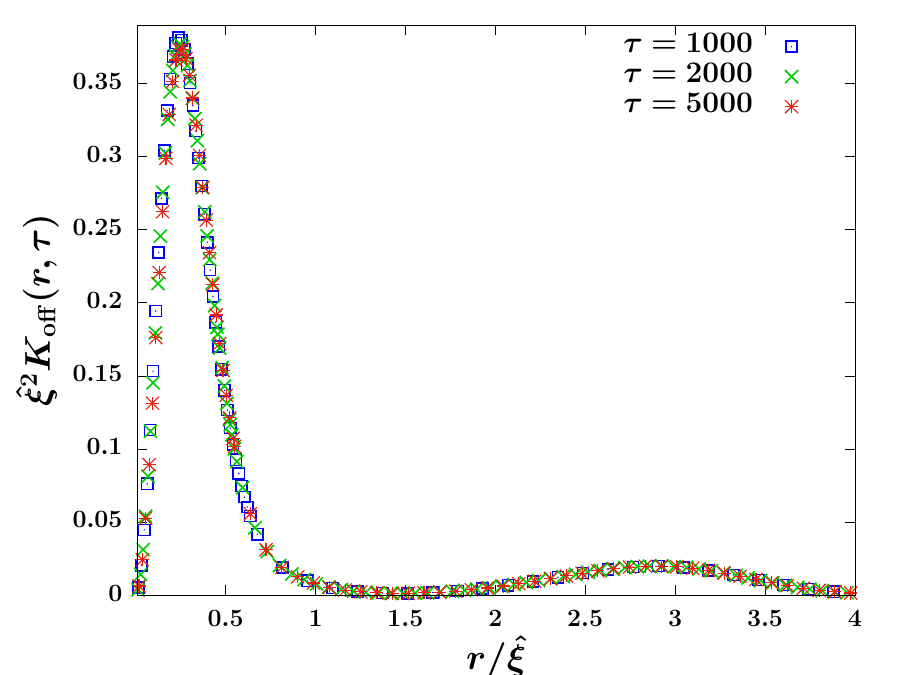}
         \caption{}
     \end{subfigure}
        \caption{Off-diagonal correlator for $g_i=10$, $g_f=0$ and $t_c=0.9\tau$ in the quench protocol (\ref{nonlinprot}) for various quench exponents. For (a) $\omega=0.75$ and (b) $\omega=1$, the inset shows that the data does not collapse using the KZ length scale, $\hat{\xi}=\tau^{\frac{\omega}{1+\omega}}$; however, as the main figure shows, the scaling function approaches the large-$\tau$ limit on using the dephasing length scale given by Eq. \eq{ellform}. 
         As shown in panel (c) for $\omega=4$, the correlator exhibits good data collapse with the KZ length scale. In all the figures, the data are obtained by numerically solving the differential equations (\ref{sm_ukpvkp}), and then integrating Eq. (\ref{koff2}).}
\label{sm_offdiag}
\end{figure} 

\clearpage 


\section{Longitudinal mean defect density}

\begin{figure}[ht]
     \centering
     \includegraphics[width=0.6\textwidth]{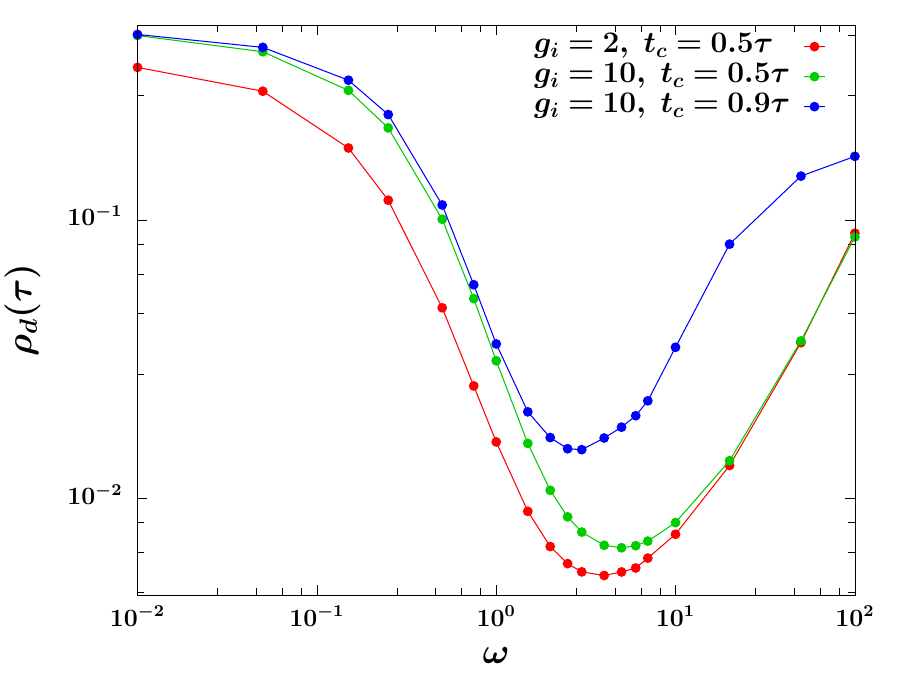}
     \caption{Longitudinal mean defect density as a function of $\omega$ for different $g_i$ and $t_c$ values and fixed $g_f=0$, $\tau=10^2$ in the quench protocol (\ref{nonlinprot}). 
     The figure shows that, for a fixed $\tau$, there exists an optimal quench exponent $\omega$ at which the mean density of defects at the end of the quench is minimum. We also note that for $\omega>1$, the final defect density depends strongly  on $t_c$ which is the time spent in the paramagnetic phase. 
          The data are obtained by numerically solving the differential equations (\ref{sm_ukpvkp}) and then integrating Eq. (\ref{rhoddef}). Here, the solid lines are guide to the eye.}
\label{SM_defden}
\end{figure}


\section{Halting time protocol}
\label{app_halt}

\begin{figure}[t!]
     \centering
     \begin{subfigure}{0.32\textwidth}
         \centering
         \includegraphics[width=\textwidth]{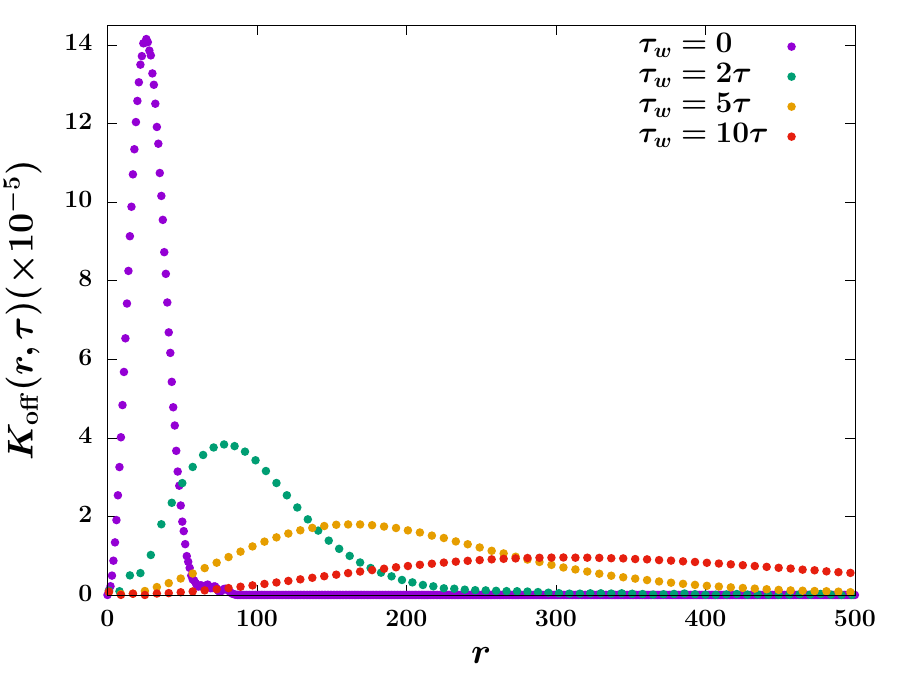}
         \caption{}
     \end{subfigure}
     \begin{subfigure}{0.32\textwidth}
         \centering
         \includegraphics[width=\textwidth]{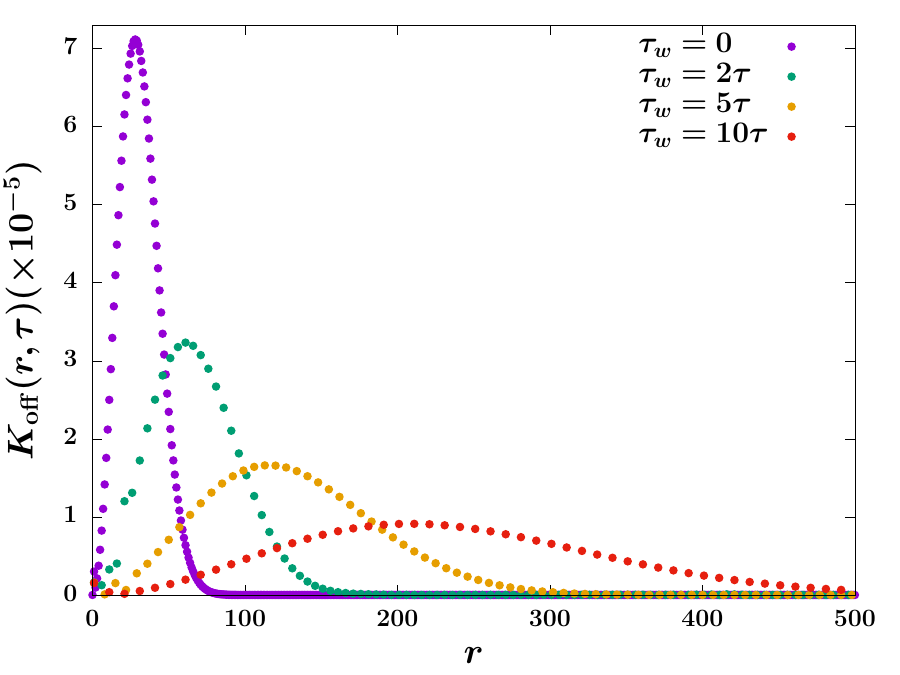}
         \caption{}
     \end{subfigure}
     \begin{subfigure}{0.32\textwidth}
         \centering
         \includegraphics[width=\textwidth]{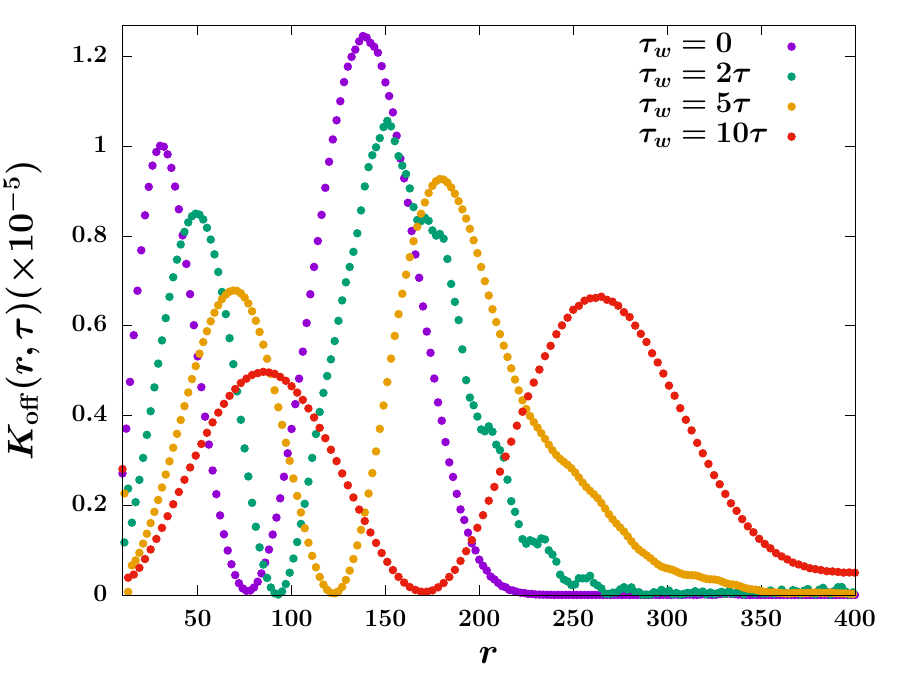}
         \caption{}
     \end{subfigure}
        \caption{Off-diagonal correlator at the end of quench as a function of distance for $g_i=2$, $g_f=0$, $\tau=10^2$ and $t_c=0.5\tau$ in the quench protocol (\ref{nonlinprot}), with different waiting times $\tau_w$ at $g_w=0.5$, for (a) $\omega=0.75$, (b) $\omega=1$, (c) $\omega=3$, respectively. The figures show that the peak height of $K_{\textrm{off}}$ decreases and the width increases with increasing waiting time $\tau_w$ indicating that the system is dephasing as the halt time is increased in the ferromagnetic phase. Additionally, for $\omega>1$, two peaks in the off-diagonal correlator are seen for any $\tau_w \geq 0$. The variation of peak height and peak location as a function of quench exponent and waiting time are shown in Fig.~\ref{tw_b}. The data are obtained by numerically solving the differential equations (\ref{sm_ukpvkp}) and then integrating Eq. (\ref{koff2}).}
\label{SM_tw}
\end{figure}

We implemented  a quench protocol with a halt by allowing the system to evolve at fixed $g_w=0.5$ for a duration $\tau_w$, and then resuming the original quench protocol. That is,
 \bsn
{g(t)=}
        g_c+(g_i-g_c) \left(\frac{t_c-t}{t_c}\right)^\omega, & $ 0 \leq t < t_c $ \\
        g_c-g_c \left(\frac{t-t_c}{\tau-t_c}\right)^\omega, & $t_c \leq t < t_w$ \\
        g_w, & $t_w\leq t \leq t_w+ \tau_w$ \\
        g_w \left(\frac{\tau+\tau_w-t}{\tau-t_w}\right)^\omega,                & $t_w+\tau_w \leq t\leq \tau+\tau_w$
\label{nonlinprot2}    
\esn
where, $g(t_w)=0.5$. The results are shown in Fig.~\ref{SM_tw}. 

\section{General equation for the nonlinear quench}
\label{app_genl}

From Eq. \eq{sm_ukvk}, it follows that for arbitrary quench protocols, $u_k, v_k$ obey the following second order differential equation with time-dependent coefficients, 
\bea
i \frac{d^2 u_k}{dt^2} &=&  -2 J \frac{d g}{dt} u_k -i (2 J)^2 [(1-g)^2+2 g (1-\cos k)] u_k \label{exactueqn} \\
i \frac{d^2 v_k}{dt^2} &=& 2 J \frac{d g}{dt} v_k -i (2 J)^2 [(1-g)^2+2 g (1-\cos k)] v_k
\eea
Then, for a protocol $g(t)=1-(\frac{t}{\tau})^\omega, -\infty < t < \tau$, in the scaling limits $k \to 0, \tau \to \infty, n=k \tau^{\frac{\omega}{1+\omega}}$ and for $x=\frac{t}{\tau}$, we obtain
\bea
\frac{i}{\tau^\frac{2}{1+\omega}} \frac{d^2 u_k}{dx^2}&=& 2 J \omega \tau^{\frac{\omega-1}{\omega+1}} x^{\omega-1} u_k  -i (2 J)^2 n^2  u_{k}- i  \tau^{\frac{2\omega}{1+\omega}} (2 J)^2 x^{2 \omega}  u_k
\eea
which, for finite $x, n$ and large $z=\tau^{\frac{1}{1+\omega}} x$ yields 
\bea
i \frac{d^2 u_k}{dz^2}&=& \left[ 2 J \omega  z^{\omega-1}   -i (2 J)^2 n^2  - i  (2 J)^2 z^{2 \omega} \right]  u_k \label{ukzeq}
\eea
 For slow quench, we are interested in the large $z$ behavior, but $z \to \infty$ is an irregular singular point of the above equation, with $u_k \stackrel{z \to \infty}{\sim} e^{C |z|^{\omega+1}}$ where, $C$ is a function of $\omega$, and is, in general, complex. 
We attempted to tackle the above equation using a WKB theory [see, for e.g., Chapter 10 of \cite{Bender:1978}], but unfortunately, except for $\omega=1$, even the integral within the geometrical optics approximation does not appear to be solvable.


\clearpage

\end{document}